\documentclass[fleqn,usenatbib]{mnras}

\usepackage{newtxtext,newtxmath}

\usepackage[T1]{fontenc}

\DeclareRobustCommand{\VAN}[3]{#2}
\let\VANthebibliography\thebibliography
\def\thebibliography{\DeclareRobustCommand{\VAN}[3]{##3}\VANthebibliography}

\usepackage{graphicx}	
\usepackage{amsmath}	
\usepackage{threeparttable}

\newcommand{\mstar}{M$_{\ast}$}
\newcommand{\msun}{M$_{\odot}$}

\title[Dwarfs in COSMOS-Web]{Multi-wavelength morphology and dust emission in low-redshift dwarf galaxies in COSMOS-Web with HST and JWST}

\author[D. Kakkad et al.]{D. Kakkad$^{1},$\thanks{E-mail: darshankakkad@gmail.com}
I. Lazar$^{1}$,
S. Harish$^{2}$,
B. Bichang'a$^{1}$,
R. K. Cochrane$^{3}$,
S. Kaviraj$^{1}$,
A. E. Watkins$^{1}$, \newauthor
G. Martin$^{4}$,
S. Koudmani$^{1,5}$,
Andrew J. Battisti$^{2,6,7}$,
Caitlin Casey$^{8,9,10}$, 
Maximilien Franco$^{11}$,\newauthor
G. Gozaliasl$^{12,13}$,
M. Hirschmann$^{14}$,
Jeyhan Kartaltepe$^{2}$,
A. A. Khostovan$^{15,2}$,
Anton Koekemoer$^{16}$, \newauthor
Daizhong Liu$^{17}$,
Henry Joy McCracken$^{18}$, 
Jason Rhodes$^{19}$, 
Brant Robertson$^{20}$
\\
$^{1}$Centre for Astrophysics Research, Department of Physics, Astronomy and Mathematics, University of Hertfordshire, Hatfield, AL10 9AB, UK\\
$^{2}$Laboratory for Multiwavelength Astrophysics, School of Physics and Astronomy, Rochester Institute of Technology, 84 Lomb Memorial Drive, \\Rochester, NY, 14623, USA\\
$^{3}$Jodrell Bank Centre for Astrophysics, University of Manchester, Oxford Road, Manchester M13 9PL, UK\\
$^{4}$School of Physics and Astronomy, University of Nottingham, University Park, Nottingham NG7 2RD, UK\\
$^{5}$St Catharine's College, University of Cambridge, Trumpington Street, Cambridge CB2 1RL, UK\\
$^{6}$Australian National University, Research School of Astronomy and Astrophysics, Canberra, ACT 2611, Australia\\
$^{7}$ARC Centre of Excellence for All Sky Astrophysics in 3 Dimensions (ASTRO 3D), Australia\\
$^{8}$The University of Texas at Austin, 2515 Speedway Blvd Stop C1400, Austin, TX 78712, USA\\
$^{9}$Department of Physics, University of California, Santa Barbara, CA 93106, USA\\ 
$^{10}$Cosmic Dawn Center (DAWN), Denmark\\
$^{11}$Université Paris-Saclay, Université Paris Cité, CEA, CNRS, AIM, 91191, Gif-sur-Yvette, France\\
$^{12}$Department of Computer Science, Aalto University, PO Box 15400,
Espoo, FI-00076, Finland\\
$^{13}$Department of Physics, University of Helsinki, P. O. Box 64, FI00014 Helsinki, Finland\\
$^{14}$Institute of Physics, Lab for galaxy evolution, EPFL, Observatoire de Sauverny, Chemin Pegasi 51, 1290 Versoix, Switzerland\\
$^{15}$Department of Physics and Astronomy, University of Kentucky, 505 Rose Street, Lexington, KY 40506, USA\\
$^{16}$Space Telescope Science Institute, 3700 San Martin Drive, Baltimore, MD 21218, USA\\
$^{17}$Purple Mountain Observatory, Chinese Academy of Sciences, 10 Yuanhua Road, Nanjing 210023, China\\
$^{18}$Institut d’Astrophsyique de Paris, UMR 7095, CNRS, UPMC Univ., Paris VI 98 bis boulevard Arago, Paris, France\\
$^{19}$Jet Propulsion Laboratory, California Institute of Technology, 4800 Oak Grove Drive, Pasadena, CA 91001, USA\\
$^{20}$Department of Astronomy and Astrophysics, University of California, Santa Cruz, 1156 High Street, Santa Cruz, CA 95064, USA\\
}

\date{Accepted XXX. Received YYY; in original form ZZZ}

\pubyear{2025}

\begin{document}
\label{firstpage}
\pagerange{\pageref{firstpage}--\pageref{lastpage}}
\maketitle

\begin{abstract}
Low-mass or dwarf galaxies (M$_{\ast}<10^{9}$ M${\odot}$) are abundant in the Universe, yet their formation and evolution remain poorly understood. Their enhanced sensitivity to feedback from star formation and active galactic nuclei (AGN) make them excellent laboratories to test whether feedback prescriptions in cosmological simulations accurately reproduce their interstellar medium (ISM) properties. We present JWST/NIRCam and MIRI imaging of nine dwarf galaxies from COSMOS-Web survey at redshift $z<0.08$, with star formation rates ranging from 0.003---0.3 M${\odot}$ yr$^{-1}$ and stellar masses of log M$_{\ast}\sim8-9$ M$_{\odot}$. The detection rate with both NIRCam and MIRI is 100\%, indicating that these dwarfs possess substantial ISM content. The detected sample includes a roughly equal mix of early-type and late-type dwarfs, suggesting that it is representative of the broader dwarf galaxy population in low-density environments. We find that the observed MIRI flux distributions are comparable to forward-modelled flux distributions of mass-matched simulated galaxies in TNG50. We further conduct a multi-wavelength morphological analysis complementing the JWST NIRCam and MIRI imaging with archival HST/ACS data, employing the CAS (concentration, asymmetry, smoothness) framework. Among the multi-wavelength images, MIRI exhibits the largest variation in CAS parameters, likely due to dust lanes and clumps in several galaxies, also suggested by Spectral Energy Distribution (SED) fitting. This suggests that the dust content in these systems may be higher than those implied by rest-frame optical or near-infrared observations alone. Upcoming UV/optical and mid-infrared spectroscopic follow-up will be critical for constraining the gas kinematics and dust grain properties of dwarf galaxies in low-density environments such as COSMOS.
\end{abstract}

\begin{keywords}
galaxies:dwarf -- galaxies:evolution -- galaxies:structure -- galaxies: ISM -- methods:observational -- techniques:image processing
\end{keywords}



\section{Introduction} \label{sect1}

Luminosity or mass functions of galaxies suggest that low-mass or dwarf galaxies are the most abundant class of galaxies in the Universe \citep[e.g.,][]{wright17}. In a hierarchical Universe, these low-mass galaxies merge and form more massive galaxies over cosmic time and therefore, studies focusing on the formation and evolution of dwarf galaxies offer valuable insights into the broader galaxy evolution process. The importance of dwarf galaxies in the Universe is further highlighted by recent works that suggest that dwarf galaxies may have contributed significantly in re-ionising the Universe at high-redshifts \citep[e.g.,][]{atek24}. Due to their low mass, they have shallow potential wells and are therefore, more vulnerable to the effects of feedback from star formation and/or Active Galactic Nuclei (AGN), compared to massive galaxies \citep[e.g.,][]{jackson21}. This makes them ideal to test implementations of current feedback prescriptions in cosmological simulations \citep[see][for a review]{vogelsberger20} by comparing the observed interstellar medium (ISM) content with predictions of low-mass galaxies in these simulations \citep[e.g.,][]{watkins25, martin25}. In fact, recent studies have indeed suggested that AGN may be more common in dwarf galaxies than previously thought \citep[e.g.,][]{kaviraj19, mezcua24} and therefore, a combination of feedback from star formation and AGN might add more energetic feedback into the ISM of dwarf galaxies \citep[e.g.,][]{dashyan18, davis22, koudmani22, partmann24, petersson25}. As a result, different feedback prescriptions (the relative strength of AGN versus star formation feedback, coupling efficiencies) may produce a different ISM composition, which can be directly tested with imaging and spectroscopic observations. 

Despite being abundant, dwarf galaxies are not targeted for high-resolution imaging or spectroscopy in as much detail as their massive counterparts, partially because of their low surface brightness. For massive galaxies, follow-up observations with even 2m-class telescopes can readily target large statistical samples. However, the low surface brightness of most dwarf galaxies poses challenges for high-resolution imaging and spectroscopic follow-up of similarly large samples. As a result, key questions surrounding the formation and evolution of these dwarfs, {such as} their star formation history, ISM content (metallicity, multi-phase gas mass), presence of outflows, morphological mix, are being actively studied. In particular, it is still unclear how the interplay between the star formation and AGN feedback shape the ISM of dwarf galaxies and to what extent does AGN affect star formation in dwarf galaxies? In other words, is the current implementation of feedback in simulations sufficient to reproduce the ISM of low-mass or dwarf galaxies? 

Most studies in the literature targeting dwarf galaxies have been biased towards nearby systems or special classes of dwarf galaxies that inherently bias samples towards higher star formation rates (SFR). Examples of such studies include targets in high-density environments, such as galaxy clusters \citep[e.g.,][]{sanchez-janssen08, ferrarese12, eigenthaler18, venhola19}, nearby dwarf galaxies \citep[e.g.,][]{dale06} or dwarfs within the Local group \citep[e.g.,][]{tolstoy09, McConnachie12, weisz14}, dwarf satellites surrounding massive galaxies \citep[e.g.,][]{trujillo21} or specific classes such as Blue Compact Dwarfs \citep[BCDs, e.g.,][]{papaderos96, wu06}. The general picture emerging from these surveys is that dwarf galaxies exhibit a wide range of morphology, from early-type to late-type and even irregular morphology. The ISM of these dwarf galaxies tends to be metal-poor, with low interstellar pressure and low dust-to-gas mass ratios \citep[see review by][]{henkel22}. Due to their low-metallicities and ongoing star formation, dwarf galaxies are also believed to be ideal laboratories for studying star formation in conditions similar to the early universe, potentially serving as low-redshift analogues to the high-redshift low-metallicity galaxies \citep[e.g.,][]{kumari19, mezcua19}.

The properties of the dwarf galaxies mentioned above are, however, studied in high-density environments which might not be representative of the broader dwarf galaxy population in the Universe. This is because dwarfs in high-density environments or nearby massive galaxies have a high probability of interactions throughout their lifetime and therefore, the host galaxy's star formation will have a high environmental dependence. In contrast, a dwarf galaxy in a low-density or ``field'' environment might experience fewer interactions and therefore might exhibit a different star formation history or metallicity compared to a typical dwarf in a high-density environment \citep[e.g.,][]{martin21}. For example, in the massive galaxy regime, \citet{peng14} showed that the lack of interactions could result in higher metallicity, especially in field galaxies. In fact, the majority of dwarf galaxies are found in low-density, or “field” environments \citep[e.g.,][]{bahcall99, dekel99, tempel12} where galaxy interactions are limited, allowing for a more secular evolutionary path \citep[e.g.,][]{martin18, jackson20, martin21}.

Studying dwarf galaxies in low-density environments requires a survey that has both a wide area as well as sufficient depth to detect targets with low surface brightness. Large-scale surveys such as the Sloan Digital Sky Survey \citep[SDSS, e.g.,][]{abazajian03} have been successful in detecting dwarf galaxies in such low-density environments \citep[e.g.,][]{bykov24}. However, SDSS is a magnitude-limited survey and is likely biased against detecting the general population of low surface brightness galaxies. Therefore, SDSS might largely capture the most actively star forming dwarf galaxies \citep[e.g.,][]{kaviraj25}. This bias is also confirmed in the spatially-resolved follow-up of dwarf galaxies via MANGA survey \citep[Mapping Nearby Galaxies at APO, e.g.,][]{bundy14}, who found that the majority of MANGA dwarf galaxies were star forming late-type galaxies and that they had significantly lower metallicities compared to massive galaxies \citep[see also][]{lequeux79,tremonti04}.

Surveys targeting deep fields, such as the Cosmic Evolution Survey \citep[COSMOS, see][]{scoville07} offer both the depth as well as wide area to detect a representative sample of dwarf galaxies in low-density environments. The ancillary multi-wavelength coverage of the COSMOS field has enabled accurate characterisation of sources within this field, including dwarf galaxies, such as star formation histories, SFR, stellar mass (M$_{\ast}$), morphological information etc. Recently, a sub-area of the COSMOS field was followed-up with JWST imaging via the COSMOS-Web survey \citep[e.g.,][]{casey23}. In this paper, we focus on multi-wavelength morphological characterisation of dwarf galaxies using imaging observations from the near-infrared Camera (NIRCam) and the mid-infrared instrument (MIRI) in a subset of dwarf galaxies in COSMOS-Web at $z<0.08$. Morphology is one of the fundamental properties of a galaxy and is believed to strongly relate to the formation history of a galaxy \citep[see][]{conselice14}. Morphological studies of dwarf galaxies have been conducted before where multiple mixes of morphological types are observed, beyond the classifications typically seen in the massive galaxy regime \citep[e.g.,][]{reaves83, sandage84}. However, as mentioned above, the vast majority of these studies target very nearby dwarfs or BCDs \citep[e.g.,][]{loose86, kunth88, vandenbergh98} and are usually limited to specific wavelength range, mostly optical.

In this paper, we select our parent sample from the low-density COSMOS field, particularly the targets in \citet{lazar24}. By combining observations across a wide range of wavelengths via HST and JWST imaging, we are able to assess the contribution of emission from dust and Polycyclic Aromatic Hydrocarbons (PAHs) in the Spectral Energy Distribution (SED) of the selected sources towards near-infrared and mid-infrared wavelengths. We also compare the observed MIRI fluxes with predicted fluxes from TNG50 suite of simulations \citep{nelson19} to infer if the star formation and ISM physics, along with feedback from star formation and/or black holes in TNG50 are sufficient to reproduce the mid-infrared properties of these dwarfs.

This paper is organised as follows: In Section \ref{sect2}, we describe the selection of the sub-sample of dwarf galaxies from COSMOS-Web. In Sect. \ref{sect3}, we briefly present the JWST and MIRI imaging observations and other ancillary data such as HST/ACS imaging used for the analysis in this paper. In Sect. \ref{sect4}, we present the analysis and results: specifically the comparison of the observed fluxes with predicted fluxes from TNG50 simulations, non-parametric morphological characterisation using CAS parameters and morphological $k$-corrections. This is followed by the discussion of results in Sect. \ref{sect5} and summary and conclusions in Sect. \ref{sect6}.\\
\indent The following cosmological parameters are used throughout this paper: $H_{0}$ = 70 km s$^{-1}$, $\Omega_{\rm M}$ = 0.3 and $\Omega_{\Lambda}$ = 0.7. North is up and East is to left in all the maps presented in this paper.

\begin{figure*}
\centering
\includegraphics[width=0.30\textwidth]{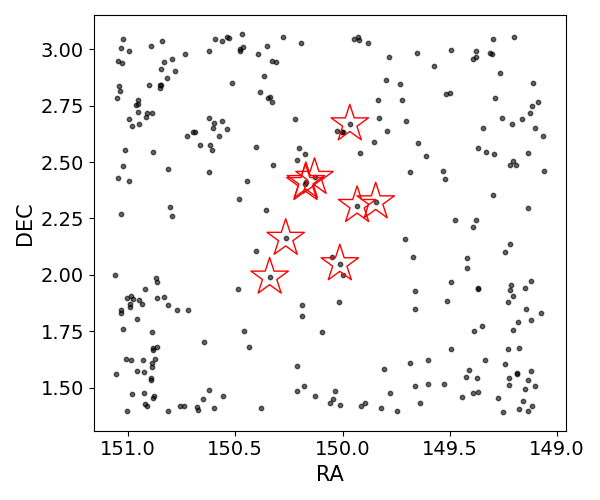}
\includegraphics[width=0.33\textwidth]{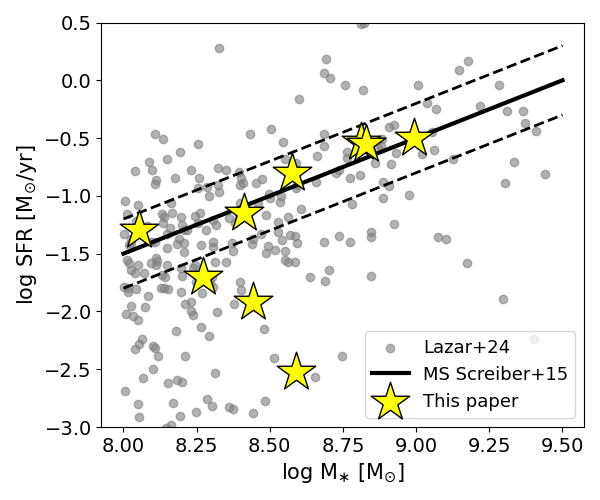}
\includegraphics[width=0.33\textwidth]{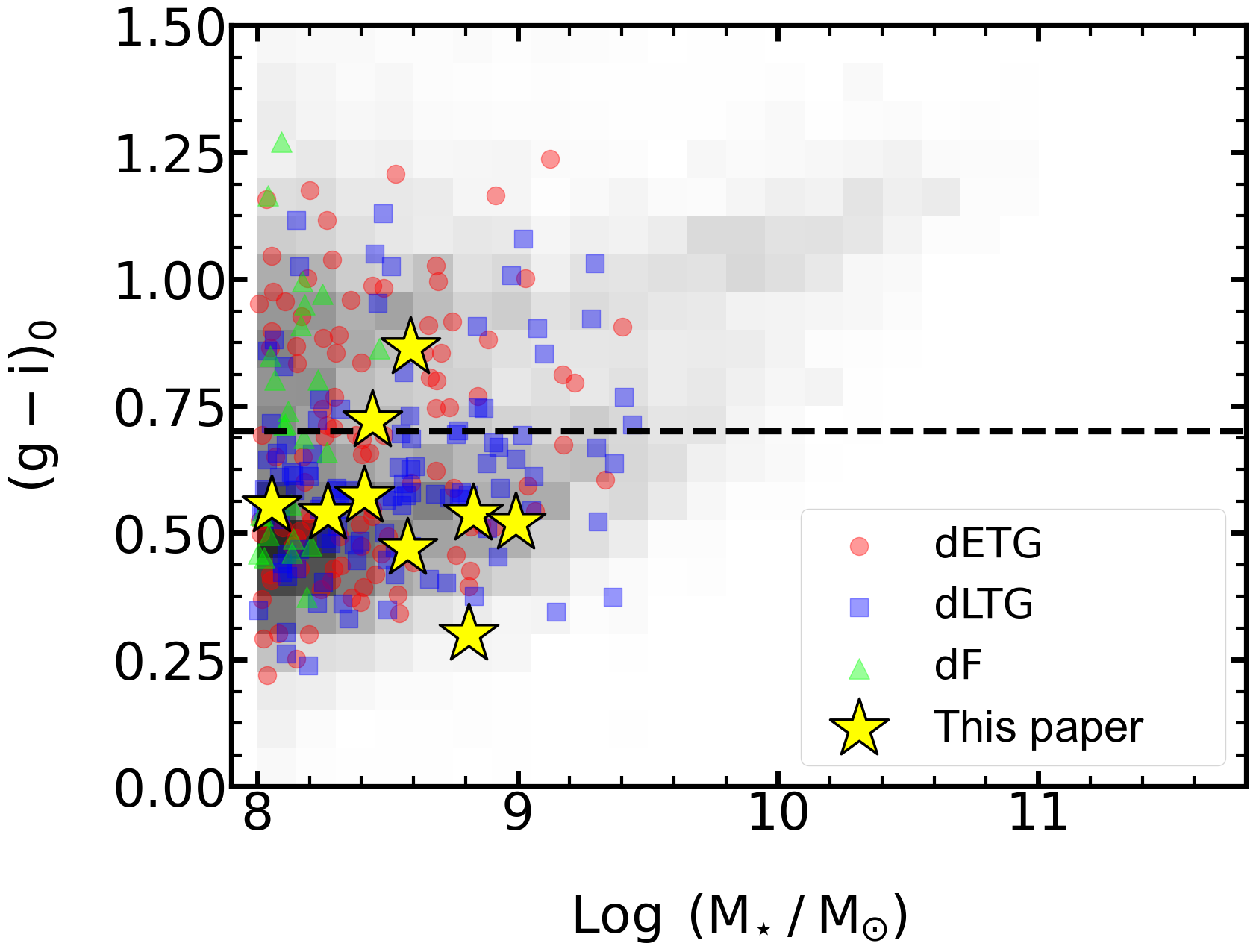}
\caption{The {\it left panel} shows the spatial distribution of the parent sample of dwarf galaxies in the COSMOS field from \citet{lazar24}. The sources highlighted by red stars are covered by NIRCam and MIRI imaging campaigns of the COSMOS-Web survey. The {\it middle panel} shows the location of the dwarf galaxies covered in the COSMOS-Web campaign, compared to the parent \citet{lazar24} sample. Most of the dwarf galaxies are located on the main sequence of massive star forming galaxies \citep[solid line with $\pm$0.3 dex regions shown by dashed line, see][]{schreiber15}, extrapolated to lower masses. Three galaxies are below the main sequence in the passive galaxy regime. The {\it right panel} shows the location of the dwarf galaxies in the colour $(g-i)_{0}$ versus M$_{\ast}$ plane. The grey colour-scale background shows the location of all galaxies from the COSMOS survey (including massive galaxies) and the coloured data points show the different morphological types of dwarf galaxies from \citet{lazar24}. The dotted line demarcates blue (bottom half) and red (top half) galaxies. The sample presented in this paper, shown as yellow stars show both red and blue dwarfs are detected with the COSMOS-Web imaging campaigns.}
\label{fig:sample_selection}
\end{figure*}

\section{Dwarf galaxies in the COSMOS survey} \label{sect2}

The parent sample of dwarf galaxies has been obtained from the COSMOS2020 catalogue \citep[see][]{weaver22}. The COSMOS field has ancillary photometric data across a wide range of wavelengths, from the UV through mid-infrared wavelengths using the following instruments: Galaxy Evolution explorer \citep[GALEX, see][]{zamojski07}, MegaCam at the Canadian France Hawaii Telescope \citep[CFHT, see][]{sawicki19}, the Advanced Camera for Surveys at the Hubble Space Telescope \citep[HST/ACS, ][]{koekemoer07}, Hyper-Suprime Cam \citep[HSC, see][]{aihara18} and Suprime-Cam \citep{taniguchi15} at Subaru Telescope, VISTA InfraRed CAMera (VIRCAM) at the Visible and Infrared Survey Telescope for Astronomy \citep[VISTA, see][]{mccracken12} and the Infrared Array Camera (IRAC) aboard the Spitzer telescope \citep[see][]{ashby18}. The multi-wavelength photometry has resulted in an accurate derivation of several host galaxy parameters such as redshift ($z$), stellar mass (M$_{\ast}$) and star formation rates (SFR) for $\sim$1.7 million sources in the COSMOS field. Further details about the COSMOS2020 catalogue is available in \citet{weaver22}.

Recently, \citet{lazar24} performed a systematic morphological characterisation of dwarf galaxies in the COSMOS2020 catalogue using optical $gri$ colour composite images from the Hyper Suprime-Cam Subaru Strategic Program (HSC-SSP) ultra-deep layer \citep[e.g.,][]{aihara18, aihara19}. The HSC imaging has the advantage that it is $\sim$5 mag deeper than the standard depth SDSS imaging and $\sim$10 mag deeper than the detection limit of the SDSS spectroscopic main galaxy sample. \citet{lazar24} used a sub-sample of 257 dwarf galaxies, which have a stellar mass range of $10^{8} < {\rm M_{\ast}} < 10^{9}$ M$_{\odot}$, and $z<0.08$. The $z<0.08$ target selection ensured sufficient spatial resolution to morphologically characterise these galaxies. Given the completeness of the COSMOS2020 catalogue for galaxies with M$_{\ast} > 10^{8}$ M$_{\odot}$ out to $z\sim 0.3$, the 257 dwarf galaxies present an unbiased, statistical and a representative benchmark sample of nearby dwarfs in low-density environments. 

We briefly describe the results in \citet{lazar24} and refer to reader to the paper for further details. 90\% of the dwarf galaxies exhibit the canonical early-type and late-type morphology that are also observed in massive galaxies. Additionally, 10\% of the dwarf galaxies also exhibit ``featureless" morphological class (also referred to as dwarf spheroidals in \citet{kormendy13}) characterised by a lack of both central light concentration (that is seen in early-types) and any structure indicative of a disk. The featureless class is not observed in the massive galaxy regime. Additionally, the structural and photometric properties of the dwarf and massive early-type galaxies diverge significantly, which suggests different formation channels of the two classes. While massive early-types and late-types are predominantly red and blue respectively, dwarf early-types and late-types share similar colour distributions. Finally, while massive early-types typically exhibit negative/flat colour gradients, dwarf early-types show positive colour gradients with $\sim$50\% of dwarf early-types showing prominent blue cores extending to $\sim$1.5 R$_e$ \citep[e.g.,][]{lazar24b}. This may suggest that the general population of dwarf galaxies may still have ongoing star formation and that their ISM may consist of significant amounts of dust and gas enriched with chemical elements. 

Given a detailed morphological characterisation, coupled with multi-wavelength photometric information from the COSMOS2020 catalogue, we derive the parent sample of dwarf galaxies from the 257 targets in \citet{lazar24}. Specifically, we search for the dwarf galaxies that are covered as a part of the near-infrared and mid-infrared imaging campaign from COSMOS-Web \citep[see][]{casey23} to characterise the warm and hot dust properties of the dwarf galaxies and compare this with archival rest-frame optical images from the HST. The JWST and HST imaging observations and data are presented in the following section. 

\begin{table*}
\centering
\begin{tabular}{lccccccc}
\hline
Target ID & RA & DEC & $z$ & log SFR & log \mstar & $(g-i)_{0}$ & Morphology\\
& J2000 & J2000 & & \msun/yr & \msun & \\
(a) & (b) & (c) & (d) & (e) & (f) & (g) & (h)\\
\hline\hline
5 & 10:00:41.546 & $+$02:24:15.080 & 0.0219 & -1.92 & 8.44 & 0.721 & E\\
61 & 10:00:41.073 & $+$02:24:46.760 & 0.0522 & -1.70 & 8.27 & 0.538 & S\\
72 & 09:59:43.716 & $+$02:18:24.630 & 0.0582 & -1.30 & 8.05 & 0.553 & S\\
104 & 10:01:21.578 & $+$01:59:21.820 & 0.0632 & -0.80 & 8.57 & 0.469 & S0\\
120 & 10:01:03.608 & $+$02:09:40.780 & 0.0637 & -0.54 & 8.81 & 0.300 & S\\
132 & 09:59:51.921 & $+$02:40:07.180 & 0.0643 & -0.55 & 8.83 & 0.536 & E\\
175 & 09:59:22.925 & $+$02:19:22.960 & 0.0683 & -1.15 & 8.41 & 0.572 & E\\
232 & 10:00:31.370 & $+$02:25:53.010 & 0.0732 & -0.50 & 8.99 & 0.520 & S\\
269 & 10:00:03.070 & $+$02:02:54.370 & 0.0782 & -2.52 & 8.59 & 0.864 & E\\
\hline
\end{tabular}
\caption{Basic properties of the dwarf sample presented in this paper. (a) Target IDs correspond to the catalog published in \citet{lazar24}, (b) \& (c) report the coordinates of the target, (d) reports the redshift, (e) and (f) report the star formation rate and stellar mass calculated from the ancillary multi-wavelength data and presented in the COSMOS2020 catalog \citep[see][]{weaver22}. Finally, (g) and (h) report the colour and morphological type of the dwarf galaxies, as calculated in \citet{lazar24} using HSC imaging campaign.}
\label{tab:sample_selection}
\end{table*}

\begin{figure*}
\centering
\includegraphics[width=0.8\textwidth]{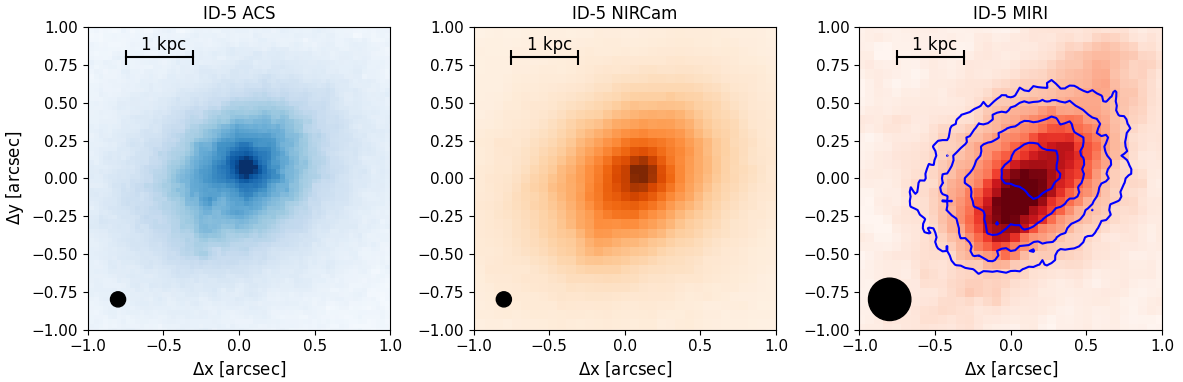}
\includegraphics[width=0.8\textwidth]{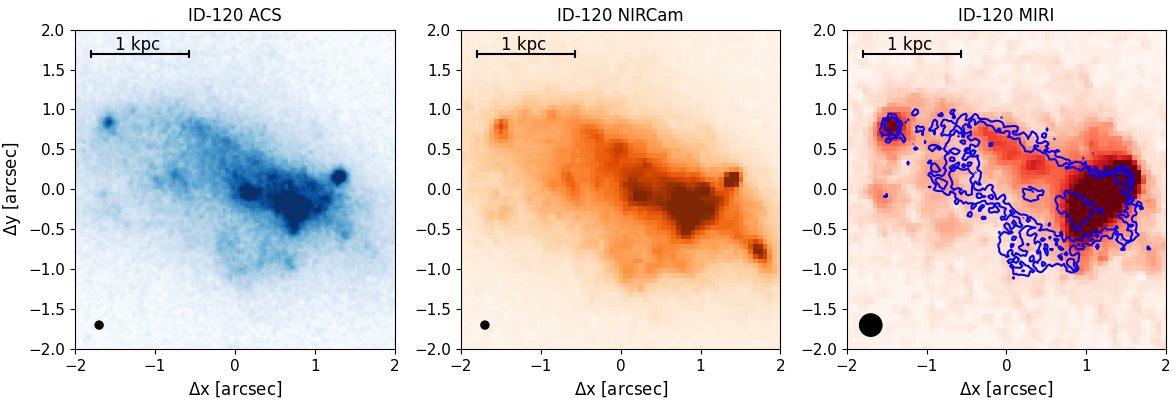}
\includegraphics[width=0.8\textwidth]{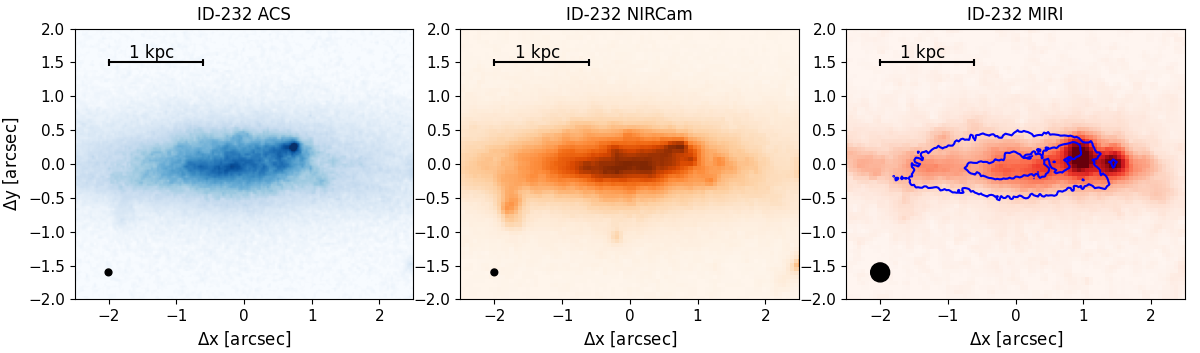}
\caption{The left, middle and right panels show the HST/ACS F814W, JWST/NIRCam F277W and JWST/MIRI F770W images of three of the dwarf galaxies (from top to bottom: IDs 5, 120 and 232) presented in this paper. The blue contours on the MIRI images in the right panel represent the F814W emission from the HST/ACS images in the left panel. The black horizontal bar shows the physical 1 kpc scale and the black circle in the bottom left shows the PSF (FWHM) of the respective images. The potential emission from the dust becomes clearly visible in the MIRI images, which are otherwise absent in the HST images. See Sect. \ref{sect4.1} for further details. }
\label{fig:multiband_images}
\end{figure*}

\section{JWST and HST observations} \label{sect3}

The COSMOS field has been observed with high-resolution imaging with HST and JWST. Here, we briefly describe these observations and the dwarf galaxy sample that are covered as a part of these surveys. 

\subsection{JWST/NIRCam and MIRI imaging} \label{sect3.1}

COSMOS-Web survey is a JWST Cycle-1 treasury programme \citep[GO-1727, PI: Kartaltepe \& Casey, see][]{casey23} that covered 0.54 $\mathrm{deg}^2$ of the COSMOS survey \citep[e.g.,][]{scoville07} with four filters of the Near-InfraRed Camera \citep[F115W, F150W, F277W, F444W, see][]{rieke23} and and 0.19 $\mathrm{deg}^2$ of the COSMOS survey with the F770W filter of the Mid-InfraRed Instrument \citep[MIRI][]{glasse15, wright23}. Here, we present a brief overview of the COSMOS-Web observations and refer the reader to \citet{casey23} for further details.

The COSMOS-Web images were obtained in multiple visits between January 2023-January 2024, with a total exposure time of $\sim$255 hours split between the NIRCam and MIRI observations. The data is reduced with the JWST calibration pipeline \citep[see][]{bushouse23_jwst1100} with further customised steps, which are elaborated in \citet{franco25} and \citet{harish25}. The final reduced NIRCam and MIRI images from the COSMOS-Web have a spatial sampling of 60 mas/pixel. For NIRCam, a finer sampling of 30 mas/pixel is available, however, we use the 60 mas/pixel sampling to compare with the MIRI images. Astrometric accuracy for both NIRCam and MIRI imaging was ensured through customised procedures detailed in \citet{franco25} and \citet{harish25}, respectively. For the NIRCam data, we employed the external JWST/HST alignment tool \citep[JHAT, see][]{rest23}, which provides improved precision compared to the current JWST pipeline adjustments. For the MIRI data, the astrometry was refined by cross-matching the MIRI source catalog with a reference catalog constructed from the COSMOS2020 survey. The NIRCam and MIRI images reach a 5$\sigma$ point-source depth of $\sim$27.5--28.2 mag and $\sim$25.3--26.0 mag, respectively. The depth of a particular pointing depends on the number of exposures at that location in the NIRCam and/or MIRI mosaic. The depth from JWST imaging campaign of the COSMOS survey is at least 1-3 magnitude deeper than the HST/ACS I-band images \citep[see][]{scoville07}.

We searched for available COSMOS-Web NIRCam and MIRI imaging coverage of the 257 dwarf galaxies from \citet{lazar24}. Unlike the NIRCam mosaic, the MIRI footprint in the COSMOS-Web survey is not contiguous, but rather it is distributed into 152 distinct regions, each region covering $4.2\times4.2$ arcmin$^{2}$, corresponding to the MIRI field-of-view. The distribution of all the 257 dwarfs in the RA-Dec plane is shown as black circles in Fig. \ref{fig:sample_selection}. Targets covered by both NIRCam and MIRI observations are highlighted as red stars. Out of the 257 dwarf galaxies in \citet{lazar24}, we find nine galaxies that are covered by the NIRCam and MIRI imaging campaigns, which are the focus of this paper. The basic properties of these nine dwarf galaxies are reported in Table \ref{tab:sample_selection}. For NIRCam, we use the F277W filter ($\lambda_{\rm pivot} = 2.776 ~\mu$m, bandwidth $=0.672 ~\mu$m) for the analysis presented in this paper, as F277W offers the highest depth compared to the other filters and has comparable spatial resolution between the different filters. For MIRI, we use the default F770W filter ($\lambda_{\rm pivot} = 7.639 ~\mu$m, bandwidth $=1.95 ~\mu$m)

The selected dwarf galaxies are in the redshift range $z\sim0.02-0.08$, have a stellar mass M$_{\ast}\sim1.1-10\times10^{8}\,$M$_{\odot}$ and star formation rates (SFR) in the range $0.003-0.3\,$M$_{\odot}\,$yr$^{-1}$. The SFRs and M$_{\ast}$ values place these dwarf galaxies on or below the main sequence of massive star forming galaxies (extrapolated to lower masses), as shown in the middle panel in Fig. \ref{fig:sample_selection}. This is also reflected in the colour-mix of the dwarf galaxies: their rest-frame $(g-i)_{0}$ values are in the range $0.3-0.9$. The right panel in Fig. \ref{fig:sample_selection} shows the distribution of the selected dwarf galaxies in the $(g-i)_{0}$ versus M$_{\ast}$ space. The morphology of these dwarfs, as determined via deep HSC imaging in \citet{lazar24}, are roughly evenly split between early-type and late-type.  The selection of both star forming and passive, blue and some red dwarfs, along with the equal split in the morphology of the sample suggests that the selected dwarf galaxies for this paper are nearly representative of the overall dwarf population within the stellar mass range $10^{8}-10^{9}\,$M$_{\odot}$.

\subsection{HST/ACS imaging} \label{sect3.2}

In addition to the NIRCam and MIRI imaging, we also searched for publicly-available archival HST/ACS images as a part of the COSMOS-HST campaign \citep[see][]{koekemoer07, scoville07}. The ACS data of the COSMOS field were taken with the F814W filter (I-band, $\lambda_{0} = 8333$ \r{A}, bandwidth $=2511$ \r{A}) for an area of 1.64 deg$^{2}$. The final reduced images have a spatial sampling of 30 mas/pixel, with an average PSF (FWHM) of 0.1 arcsec. The HST images were rebinned to a pixel scale of 60 mas to be consistent with the NIRCam and MIRI images. All the nine dwarf galaxies with the NIRCam and MIRI coverage presented in Sect. \ref{sect3.1} have publicly available HST/ACS imaging as well.

\medskip
\noindent
The HST/ACS, NIRCam and MIRI images of three of the dwarf galaxies (IDs: 5, 120 and 232) are shown as examples in Fig. \ref{fig:multiband_images}. Plots for the rest of the targets are moved to the Appendix \ref{sect:appendix}. 

\section{Analysis and results} \label{sect4}

The primary goals of this paper are to draw inferences on the characteristics of the ISM of these dwarfs i.e., whether we expect the dwarfs to exhibit the level of mid-infrared emission as observed in the MIRI images and estimate the change in the morphological characteristics of the dwarf galaxies with wavelength i.e., from optical (HST) to near-infrared (NIRCam) and mid-infrared (MIRI).

To analyse the HST and JWST images described in Sect. \ref{sect3}, we first extract rectangular cut-outs from the respective image mosaics. The cut-outs have an area of $\sim 15\times 15$ arcsec$^{2}$, which roughly cover the spatial extent of the dwarf galaxies. Depending on the redshift of the targets, this covers a physical scale of $\mathbf{\sim 7\times7}$ to $\mathbf{\sim 22\times22}$ kpc$^{2}$. The spatial resolutions of the ACS/F814W, NIRCam/F277W and MIRI/F770W are $\sim$0.1, 0.1 and 0.28 arcsec, respectively. Clearly, the HST and NIRCam images offer nearly three times better spatial resolution than the MIRI images. The images shown in Fig. \ref{fig:multiband_images}  (and Figs. \ref{fig:multiband_images_1} and \ref{fig:multiband_images_2} in the Appendix \ref{sect:appendix}) show the true resolution of the final reduced images. However, for comparative morphological characterisation, presented in this section, it is important to analyse images that have similar spatial resolution. As MIRI imaging has the worst spatial resolution, we convolve the NIRCam and ACS images with a Gaussian kernel to match them to the resolution of the MIRI images.  

\subsection{NIRCam and MIRI detection rates of dwarf galaxies} \label{sect4.1}

\begin{figure*}
\centering
\includegraphics[width=0.33\textwidth]{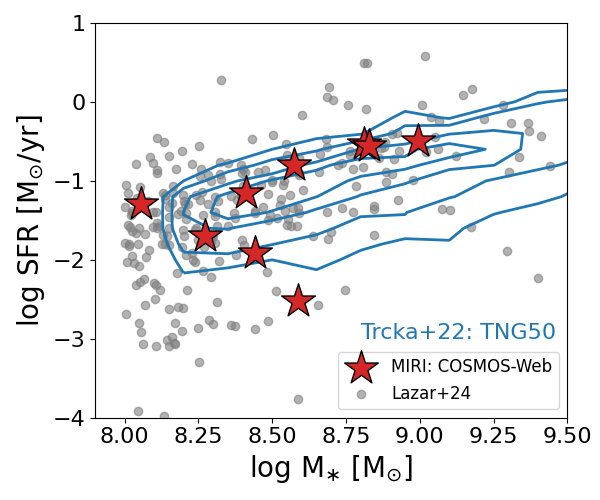}
\includegraphics[width=0.33\textwidth]{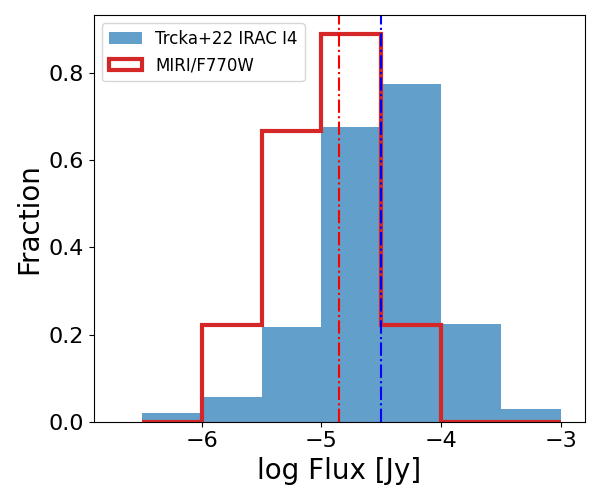}
\includegraphics[width=0.33\textwidth]{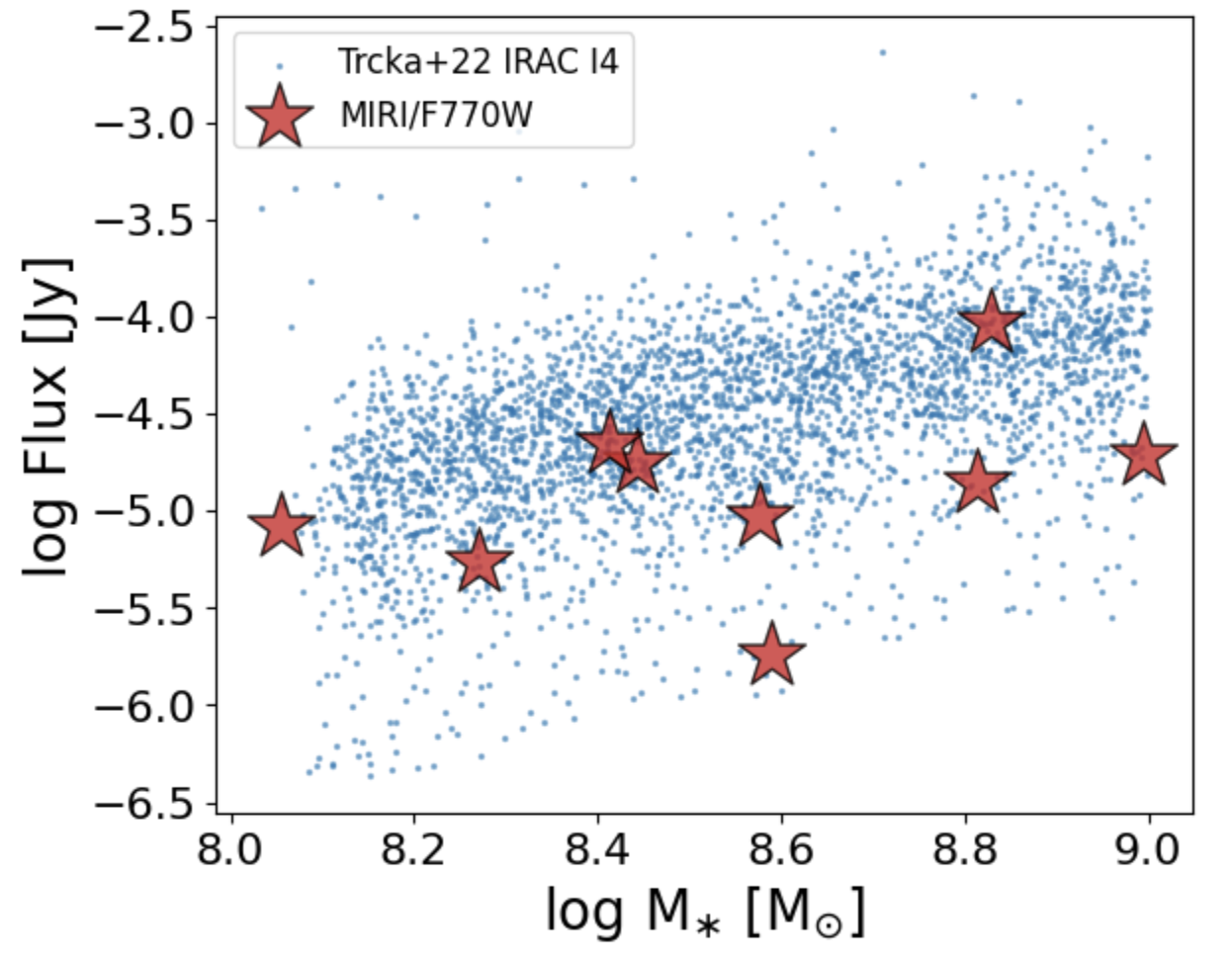}
\caption{The {\it left panel} shows the location of the TNG50 galaxies (blue contours), along with the parent dwarf galaxy sample (grey circles) and the sample  presented in this paper (red stars) in the SFR-M$_{\ast}$ plane. The {\it middle panel} shows the flux distributions of the mock IRAC-I4 flux from \citet{trcka22} (blue histogram) and the observed MIRI/F770W fluxes in this paper (red histogram outline). The {\it right panel} shows the fluxes in the middle panel as a function of stellar mass. The colour scheme in the right panel is the same as the left panel. Further details about the comparison between TNG50 data and MIRI observations are given in Sect. \ref{sect4.1}.}
\label{fig:sim_comparison}
\end{figure*}

We begin by discussing the detection rates of dwarf galaxies in the near- and mid-infrared wavelengths. All nine galaxies are detected in NIRCam and MIRI imaging from the COSMOS-Web survey, resulting in a 100\% detection rate. Given the shallow gravitational potential wells of dwarf galaxies, this raises important questions about their ISM content and composition, as well as the effectiveness of stellar and AGN feedback in expelling gas and dust from their ISM. The observed 100\% detection rate in both wavelength ranges suggests that dwarf galaxies largely retain some of their ISM, potentially in terms of their warm gas and dust content, including PAHs.

We verify if the observed fluxes in the NIRCam and MIRI images are consistent with expectations from the fluxes of low-mass galaxies in the TNG50 simulation \citep[e.g.,][]{nelson19, pillepich19}. TNG50 simulates the evolution of a $\sim$50 Mpc$^{3}$ box from cosmological initial conditions to $z = 0$ using the moving-mesh code \texttt{Arepo} \citep[e.g.,][]{springel10, weinberger20}. TNG50 has the highest resolution within the IllustrisTNG suite of simulations \citep[e.g.,][]{marinacci18, naiman18, springel18}. We compare our observed MIRI flux with mock IRAC I4 flux from \citet{trcka22}, who implemented the radiative transfer (RT) code,  Stellar Kinematics Including Radiative Transfer \citep[\texttt{Skirt}, e.g.,][]{baes11, camps15} on $\sim$14,000 $z\leq0.1$ galaxies drawn from the TNG50 simulation. The RT implementation in \citet{trcka22} modelled light from evolved stars, star-forming regions and the diffuse dust. Emission from stars was modelled for star particles with ages above $10\,$Myr using \citet{bruzual03} stellar population synthesis models, with a Chabrier initial mass function \citep{chabrier03}. Young star-forming regions were modelled with \texttt{MAPPINGS-III} spectral energy distribution (SED) templates \citep{groves08}, including a photodissociation region with a covering fraction. The diffuse dust was modelled only in relatively dense and cold gas cells according to the boundary condition criteria in \citet{torrey12} and \citet{torrey19}. Following detailed calibration by \citet{trcka22}, a fixed dust-to-metal mass ratio of 20\% was assumed. The dust grain distribution was modelled with the THEMIS dust models \citep{Jones2017} using $15$ grain size bins. Further details of the RT implementation are available in \citet{trcka22}. 

Forward-modelled fluxes in \citet{trcka22} are available for several broad-band filters from UV to sub-mm wavelengths. For the purpose of this paper, we choose the mock IRAC I4 fluxes, as the MIRI/F770W filter used in this paper is the closest match to the IRAC-I4 band ($\lambda_{0}^{\rm IRAC-I4}\sim7.8\mu$m). We use the SNAPSHOT\_91 data, corresponding to a redshift of $z\sim0.1$, which is also closest to the median redshift ($z\sim0.08$) of the observed dwarf sample presented in this paper. Figure \ref{fig:sim_comparison} shows the comparison between the observed dwarf sample presented here and the TNG50 sample. The parent dwarf galaxy sample (grey circles) and the sample selected here (red stars) lie within the area occupied by the TNG50 galaxies (blue contours) in the SFR-M$_{\ast}$ plane at log M$_{\ast}$/ M$_{\odot}< 9$, as shown in the left panel in Fig. \ref{fig:sim_comparison}. This further confirms the representative nature of the dwarf galaxies selected in this paper. The middle panel in Fig. \ref{fig:sim_comparison} shows the mid-infrared flux distributions of the observed MIRI sample (red histogram) and the IRAC-I4 TNG50 sample (blue histogram). The median values of the flux distributions are within 0.5 dex of each other, as apparent from the vertical red and blue lines in the middle panel in Fig. \ref{fig:sim_comparison}. We performed a Kolmogorov-Smirnov (KS) test \citep{hodges58} to test whether the underlying observed and simulated flux distributions are different. For a null hypothesis that the samples are drawn from two different distributions, the KS test yields a $p$-value of 0.4, which is insignificant at the standard 5\% level. Therefore, we conclude that the observed and the simulated flux distributions are similar. In other words, the simulated TNG50 mid-infrared fluxes are in agreement with the observed mid-infrared fluxes from MIRI. The quoted significance levels come with a few caveats. First, the computed $p$-value is sensitive to the choice of binning in the histogram, which we set according to the maximum uncertainty expected in flux values ($\sim 0.5$ dex) in both the simulated and observed data. In addition, the significance is influenced by the limited number of observed galaxies ($<10$), sharply in contrast to the much larger simulated sample ($\approx$14,000), leading to low-number statistical effects on the observational side. Lastly, the right panel in Fig. \ref{fig:sim_comparison} shows the mid-infrared flux as a function of stellar mass for the MIRI (red stars) and TNG50 (blue circles) sample. We do not see any systematic offsets between the observed and simulated fluxes with mass. This suggests again that the simulations are roughly in agreement with the observations for the mass range probed in this paper. We discuss this comparison between observations and simulations further in Sect. \ref{sect5}. 

\subsection{Multi-wavelength morphological characterisation} \label{sect4.2}

\subsubsection{Visual characterisation} \label{sect4.2.1}

Fig. \ref{fig:multiband_images} shows the HST/F814W (I-band, left panel), NIRCam/F277W (middle panel) and MIRI/F770W (right panel) images of three dwarf galaxies presented in this paper, as examples. Plots for the rest of the sample are moved to the Appendix \ref{sect:appendix} (Figs. \ref{fig:multiband_images_1} and \ref{fig:multiband_images_2}). The blue contours on the MIRI images (right panel in these figures) show the HST/F814W emission shown in the left panels. From these images, it is clear that the nine dwarf galaxies detected with JWST imaging have a wide range of morphology: from early-type to late-type, including mergers. In this section, we discuss the morphological characteristics of these dwarf galaxies. 

\begin{figure}
\centering
\includegraphics[width=0.95\linewidth]{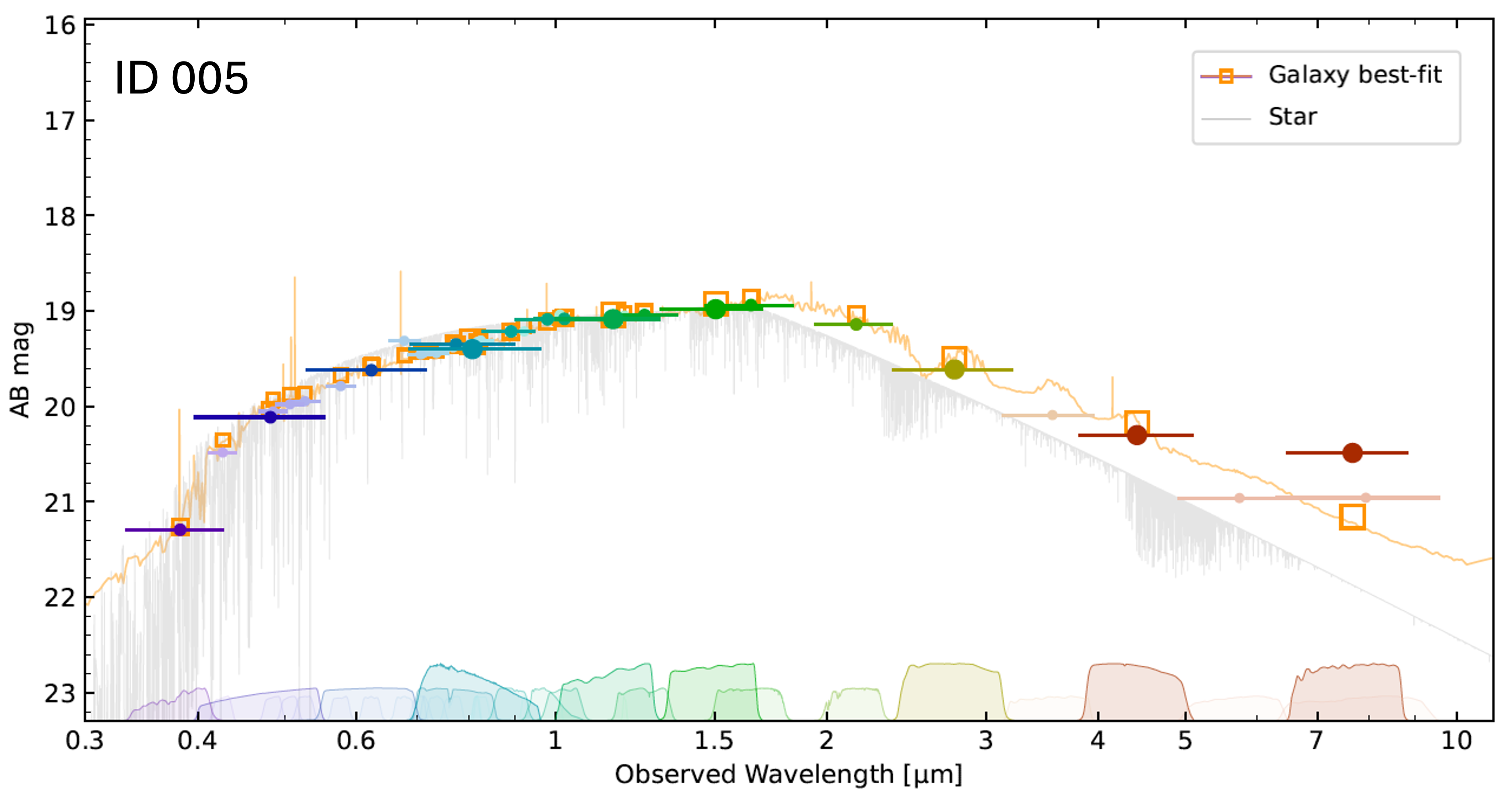}\\
\includegraphics[width=0.95\linewidth]{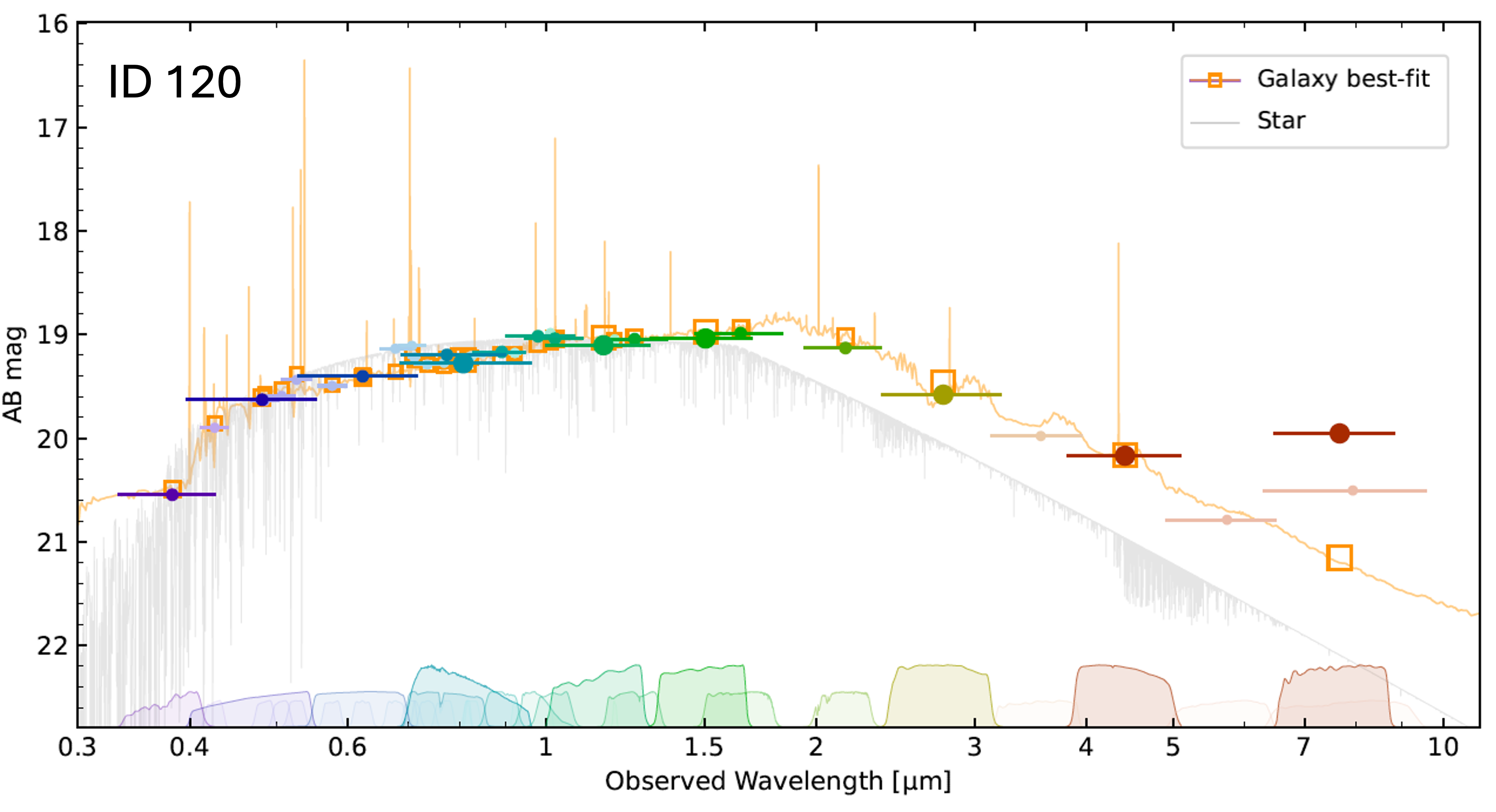}\\
\includegraphics[width=0.95\linewidth]{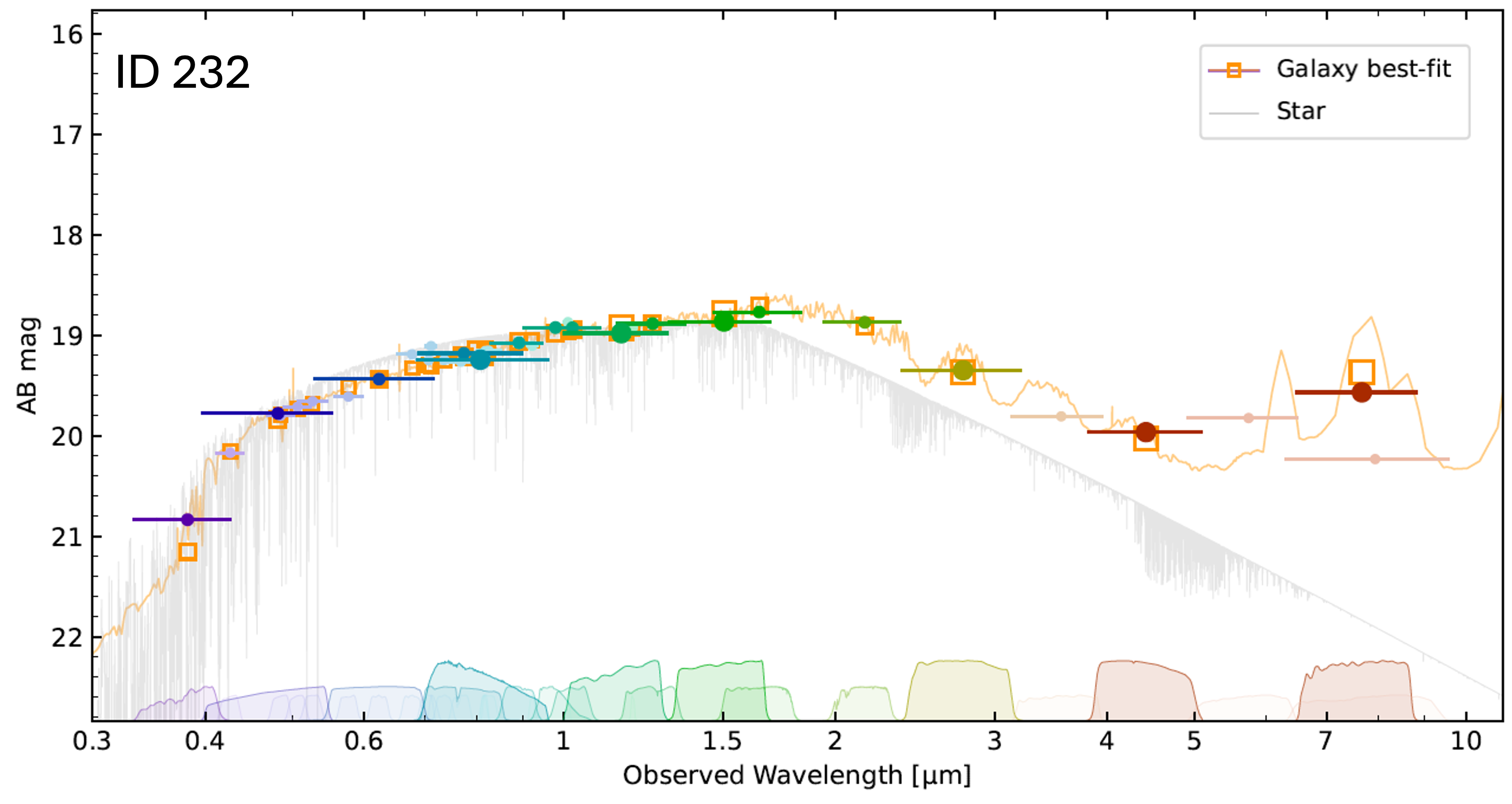}
\caption{SED fits of targets ID-5 (top panel) and ID-120 (middle panel) and ID232 (bottom panel) using \texttt{PROSPECTOR}. The coloured circles show the different photometric data points from COSMOS2025 \citep{shuntov25} corresponding to the following instruments and filters: HSC ($g,r,i,z,y$), UVISTA ($Y, J, H, K$), IRAC (1, 3, 4), HST (F814W), NIRCam (F115W, F150W, F277W, F444W) and MIRI (F770W). The orange curve shows the best-fit SED model. In all the targets, contribution from small and large dust grains and PAHs becomes apparent at mid-infrared wavelengths, the excess emission above stellar model (black body fit shown in grey colour). This is further re-inforced by the multi-wavelength visual analysis of these targets, that already indicated possible presence of dust lanes and dust clumps.}
\label{fig:sed_fits}
\end{figure}

Based on visual classification, the MIRI detected sample is evenly split between early-types and late-types, indicating that dwarf early-types retain some of their ISM like their late-type counterparts. The distribution of the dwarfs into these two categories is consistent with the rest-frame $(g-i)_{0}$ colours of the dwarf galaxies reported in \citet{lazar24}. In fact, looking at the colour distribution of the nine dwarf galaxies (see right panel in Fig. \ref{fig:sample_selection}), two of the dwarf galaxies show red colours. This corresponds to $\approx$20\% red fraction, although these are low number statistics. The observed red fraction in JWST sources is consistent with the red fraction in the parent population of dwarf galaxies \citep[13--40\%][]{lazar24}, albeit this number has large uncertainties. We find two interacting dwarfs in our sample (ID-072 and ID-120) i.e., 20\% of the sample, which is within the range found in the parent sample of dwarf galaxies in \citet{lazar24}. 

\begin{table*}
\centering
\begin{tabular}{l|cccc|cccc|}
ID & \multicolumn{3}{c}{HSC} & \multicolumn{3}{c}{HST}\\
\hline
& $n$ & C & A & S & $n$ & C & A & S\\
\hline\hline
5 & 1.7 & 2.96 & 0.03 & -0.001 & 1.0 & 2.90 & 0.03 & 0.004\\
61 & 1.4 & 3.07 & 0.14 & 0.02 & 1.0 & 2.54 & -0.02 & -0.03\\
72 & 0.7 & 2.54 & 0.24 & 0.03 & 0.6 & 2.22 & 0.23 & 0.02\\
104 & 1.9 & 3.34 & 0.03 & 0.00 & 1.1 & 2.92 & 0.03 & 0.01\\
120 & 0.8 & 2.89 & 0.20 & 0.04 & 0.7 & 2.26 & 0.30 & 0.02\\
132 & 1.9 & 3.09 & 0.06 & 0.085 & 1.1 & 2.86 & 0.09 & 0.004\\
175 & 1.2 & 2.83 & 0.02 & 0.08 & 0.9 & 2.74 & 0.02 & 0.02\\
232 & 1.1 & 2.69 & 0.10 & 0.04 & 0.8 & 2.66 & 0.11 & 0.02\\
269 & 2.0 & 3.10 & 0.01 & 0.05 & 1.2 & 3.07 & 0.03 & -0.003\\
\hline
\end{tabular}
\caption{Parametric and non-parametric morphological analysis of the dwarf galaxies presented in this paper. The table compares the S\'{e}rsic index ($n$) and `CAS' parameters (C = Concentration, A = Asymmetry, S = Clumpiness) from HSC/I-band \citep{lazar24} and HST/I-band images shown in Figs. \ref{fig:multiband_images}, \ref{fig:multiband_images_1} and \ref{fig:multiband_images_2}. See Sect. \ref{sect4} for further details.}
\label{tab:cas_hsc_hst}
\end{table*}

\begin{figure*}
\includegraphics[width=0.8\textwidth]{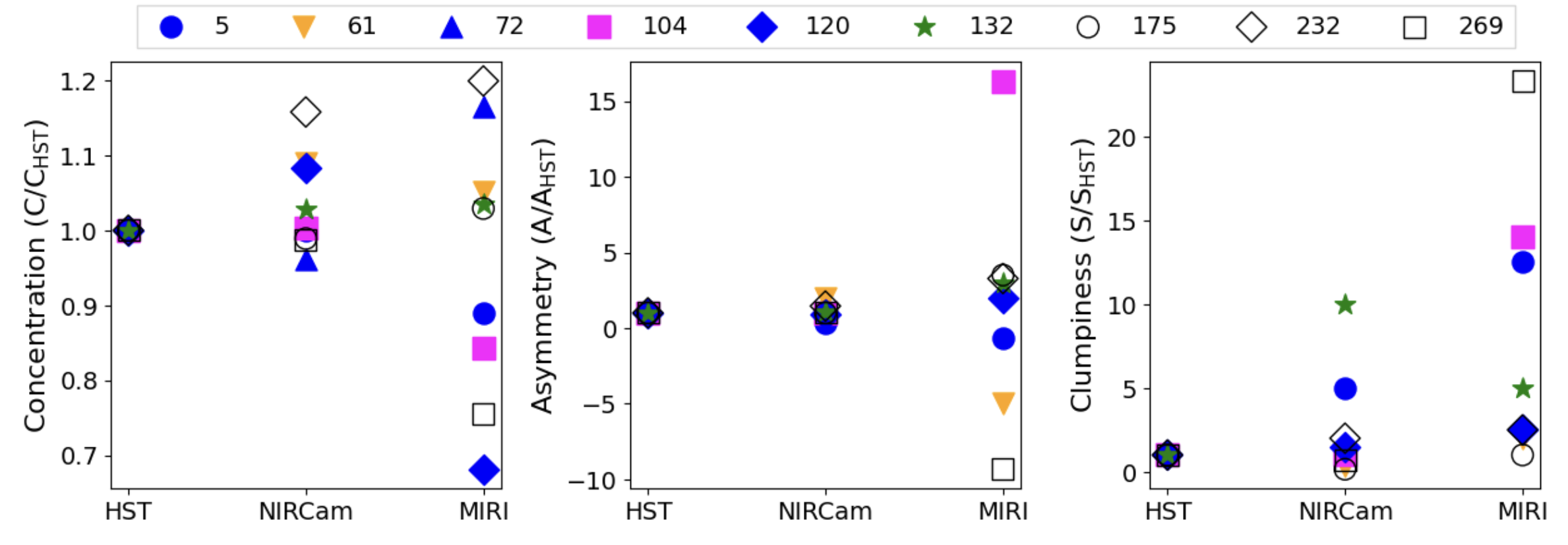} 
\caption{The {\it top panel} shows the Concentration (C, left panel), asymmetry (A, middle panel) and Clumpiness (S, right panel) as a function of the HST/F814W, NIRCam/F277W and MIRI/F770W for the individual targets (different colours and symbols) presented in this paper. The values in each filter have been normalised to the HST/F814W to guide the change in the respective parameters.}
\label{fig:CAS_HST_NIRCam_MIRI}
\end{figure*}

\begin{figure*}
\includegraphics[width=0.8\textwidth]{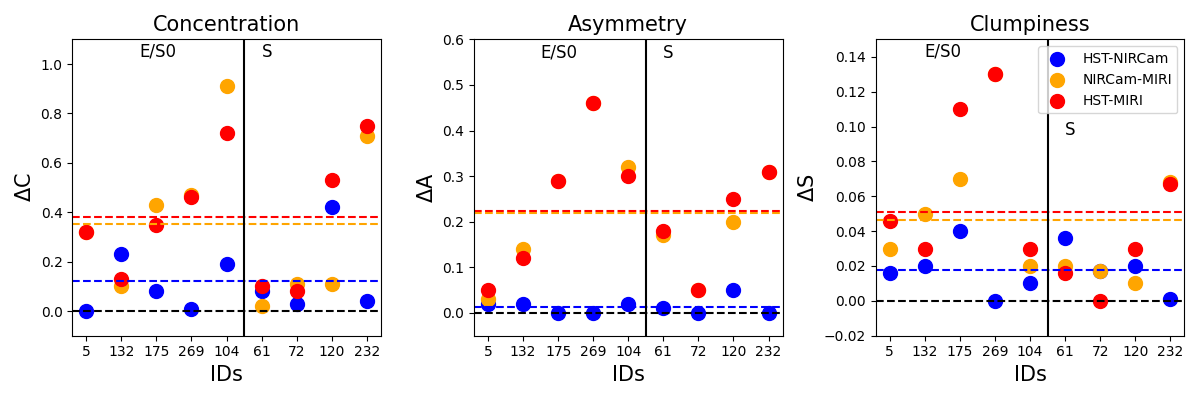}
\includegraphics[width=0.8\textwidth]{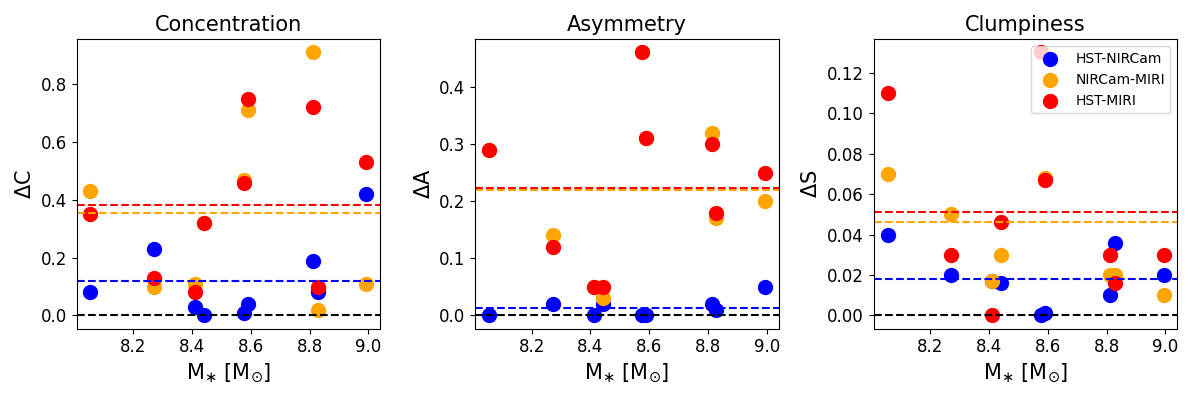}  
\caption{Change in the concentration (C, left panels), asymmetry (A, middle panels) and clumpiness (right panels) as a function of the morphological type in the top panels: early-type (E/S0) or late-type (S) and as a function of stellar mass in the bottom panels. Blue colour shows the difference in the CAS parameters between the HST and NIRCam images, orange shows the difference between NIRCam and MIRI images and red shows the difference between HST and MIRI images. The black dotted line in all panels show the $\Delta {\rm CAS} = 0$ line and the blue, orange and red dotted line shows the average change for the HST-NIRCam, NIRCam-MIRI and HST-MIRI images, respectively. Largest differences in the parameters are seen between HST or NIRCam versus MIRI images.}
\label{fig:DeltaCAS_plots}
\end{figure*}

\begin{figure}
\includegraphics[width=0.45\textwidth]{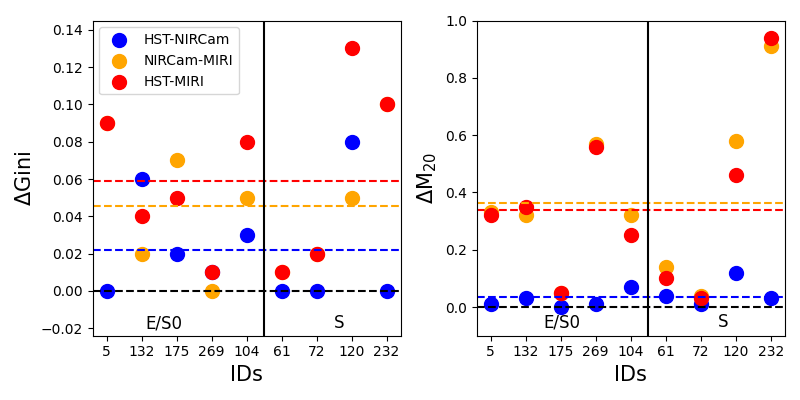}
\includegraphics[width=0.45\textwidth]{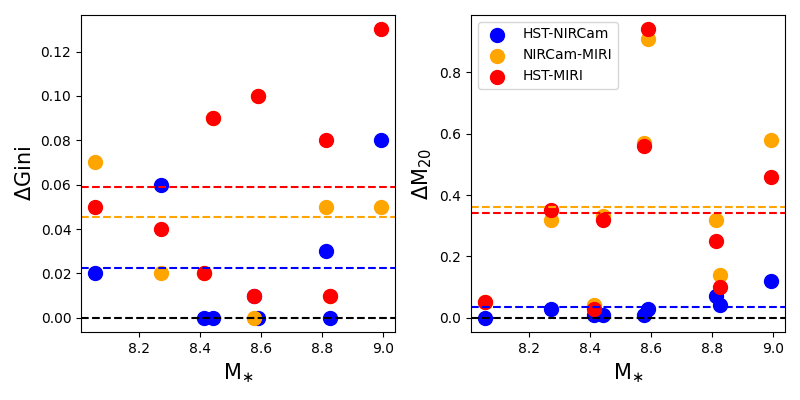} 
\caption{The figures show the change in  Gini ($\Delta$Gini, left panels) and M$_{20}$ ($\Delta$M$_{20}$, right panels) with different datasets as a function of the morphological type (top panels) and stellar mass (bottom panels). Colour scheme of these plots is the same as Fig. \ref{fig:DeltaCAS_plots}.}
\label{fig:DeltaM20Gini_plots}
\end{figure}

\begin{figure*}
\centering
\includegraphics[width=0.9\textwidth]{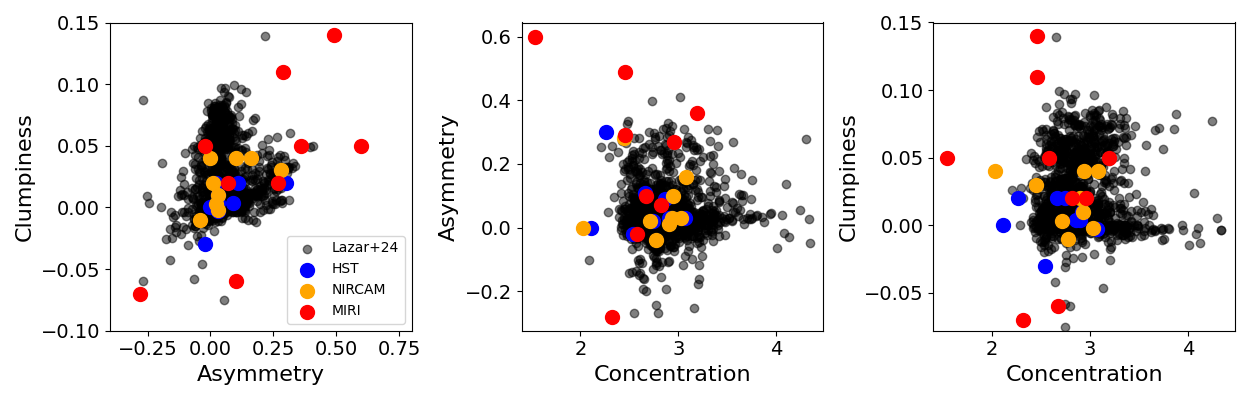}
\caption{Location of the COSMOS-Web dwarf galaxies presented in this paper, with respect to the parent \citet{lazar24} sample (background black circles) in different CAS parameter space. Blue, orange and red circles show the morphological parameters determined from the HST/ACS F814W, JWST/NIRCam F277W and JWST/MIRI F770W images, respectively. While most galaxies are within the locii of the \citet{lazar24} sample, there is a noticeable change in the parameter location with wavelength. }
\label{fig:CAS_plot}
\end{figure*}

\begin{figure}
\centering
\includegraphics[width=0.4\textwidth]{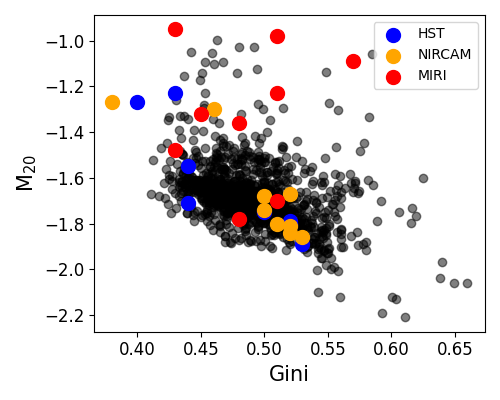}
\caption{M20 vs Gini coefficient for the COSMOS-Web dwarf galaxies presented in this paper with respect to the parent \citet{lazar24} sample. Colour scheme is the same as in Fig. \ref{fig:CAS_plot}. Similar to Fig. \ref{fig:CAS_plot}, the COSMOS-Web dwarfs approximately occupy the same locations as the parent sample. }
\label{fig:M20_vs_Gini}
\end{figure}

In summary, despite the low number statistics presented due to the coverage of MIRI imaging, based on the visual classifications and the colour distribution of dwarf galaxies in our sample, our results suggest that the MIRI detections are not strongly biased towards a particular class or colour of dwarf galaxies. The selected dwarfs span a wide range of the SFR-M$_{\ast}$ plane, as shown in Fig. \ref{fig:sim_comparison}, which suggests that the sub-sample of dwarf galaxies presented here is representative of larger dwarf population in low-density environments.

Although the morphological classification itself does not change as a function of the wavelength, there are subtle differences that may suggest the presence of dust lanes or dust clumps and regions subjected to dust obscuration in these galaxies. For example, in ID-5 (top panel in Fig. \ref{fig:multiband_images}), the HST image shows a lack of emission towards the SE of the nucleus. The MIRI image, on the other hand, shows enhanced emission in this region, suggesting that the rest-frame optical emission is possibly obscured by a dust lane, as seen in the Milky Way as well as many nearby massive galaxies (e.g., NGC 7172). The dust lanes observed in early-type galaxies like ID-5 may also point to past merger activity in these dwarf systems \citep[e.g.,][]{kaviraj12}. Evidence of dust clumps is also seen in ID-120, where the MIRI image reveals excess mid-infrared emission toward the western regions of the galaxy, in contrast to the lack of corresponding emission in the HST and NIRCam images.

We also obtained SED fits from the COSMOS-Web survey, as described in \citet{shuntov25}. To validate these results, we performed independent SED fitting using the \texttt{PYTHON}-based \texttt{PROSPECTOR} code \citep[see][]{johnson21, bichanga24}, which are consistent with the SED fits in \citet{shuntov25}. Figure \ref{fig:sed_fits} presents example SED fits for ID-5, ID-120, and ID-175 in the top, middle, and bottom panels, respectively. The SED fits for the rest of the sample presented in this paper are shown in Fig. \ref{fig:sed_fits_appendix}. The fits clearly show a substantial contribution from dust grains and PAHs in the mid-infrared, beyond the stellar component, which is represented by the grey blackbody model in Fig. \ref{fig:sed_fits}. For the galaxies presented in this paper, we estimate that stellar emission contributes anywhere between about 6–70\% of the rest-frame $\sim 7\mu$m flux, with a median of 45\%, indicating that for most sources, more than half of the mid-infrared emission arises from warm or hot dust and PAH features. The exact nature of the emission at mid-infrared wavelengths can be verified with dedicated spectroscopic follow-up with the medium resolution spectrograph (MRS) of MIRI. The implications of these dust features, including comparison with studies in the massive galaxy regime is further discussed in Sect. \ref{sect5}.

\subsubsection{Parametric and non-parametric morphological characterisation} \label{sect4.2.2}

In addition to the visual classification done above, parametric and non-parametric methods can be employed to quantify the morphological characteristics of the dwarf galaxies detected in different wave bands. Parametric methods such as fitting the galaxy surface brightness with S\'{e}rsic profile are useful to estimate the S\'{e}rsic index, which can then be compared against different classes (early-types or late-types) and also can be compared against the values observed in the massive galaxy regime. We have used the \texttt{STATMORPH} package to derive S\'{e}rsic parameters using two-dimensional single-component S\'{e}rsic profile fits using \texttt{PYTHONASTROPY} modelling package within \texttt{STATMORPH}. Further details on S\'{e}rsic fits are available in \citet{rodriguez-gomez19} and \citet{lazar24}. The distribution of S\'{e}rsic indices in the parent dwarf ETG and LTG from \citet{lazar24} show overlapping values with the median S\'{e}rsic index of 1.3 in late-type and 1.5 in early-type dwarf galaxies. The S\'{e}rsic indices calculated from imaging using space telescopes are even lower than that of the parent dwarf galaxies. This is clear from the values quoted in tables \ref{tab:cas_hsc_hst} and \ref{tab:cas_miri_nircam}. The average S\'{e}rsic index for dwarf early-types from HSC/I-band is 1.7, compared to 1.1 with HST/I-band. Similarly, the average S\'{e}rsic index for dwarf late-types from HSC/I-band is 1.0, compared to 0.8 with HST/I-band. Therefore, for the same waveband, ground-based imaging results in higher S\'{e}rsic indices, compared to space telescope imaging. This may be due to larger PSFs in ground-based images and/or the detection or more diffuse emission sensitive in ground-based HSC observations, which are deeper than the HST imaging. Furthermore, the S\'{e}rsic indices appear to decrease as a function of wavelength, at least for the dwarf ETGs (1.1 to 0.9 from HST to MIRI imaging), while it is nearly constant for dwarf late-type galaxies. The difference in S\'{e}rsic indices observed in dwarf galaxies is less stark when compared to ETGs and LTGs in massive galaxies, where the corresponding S\'{e}rsic indices are 3.5 and 1.5, respectively \citep[see also][]{faber83}.

Non-parametric methods, on the other hand, do not require fitting the surface brightness profiles with a known function. For example, studies by \citet{conselice03}, \citet{abraham03} and \citet{lotz04} introduced morphological parameters such as concentration (C), asymmetry (A), clumpiness (S), M$_{20}$ and the Gini coefficient. We refer the reader to the references \citep[see also][]{lazar24} for the definitions of each of these parameters. The former three are collectively known as `CAS' parameters. In the nearby Universe, such non-parametric morphological characterisation has suggested that massive ETGs show higher concentration, lower asymmetry and lower clumpiness than their LTG counterparts. We aim to perform a similar analysis for our dwarf galaxy sample in different wavebands. 

We refer the reader to \citet{lazar24} for details regarding the derivation of CAS parameters. The process is briefly explained here. Before determining the CAS parameters, we masked external sources such as interlopers, foreground stars and foreground or background galaxies that would otherwise have interfered with the surface brightness profile of the primary dwarf sources. The asymmetry, clumpiness, concentration, M$_{20}$ and Gini coefficient were estimated using the \texttt{statmorph} package in \texttt{python} \citep[see][]{rodriguez-gomez19}. These parameters derived from the HST and JWST images, including ground-based HSC imaging are reported in Tables \ref{tab:cas_hsc_hst} and \ref{tab:cas_miri_nircam}.

Similar to previous studies targeting well-resolved galaxies in the nearby Universe, we see that dwarf early-types exhibit a higher concentration, lower asymmetry, and lower clumpiness than their late-type counterparts across all the HST, NIRCam and MIRI images \citep[see results from massive galaxies in][]{conselice03, holwerda14, cheng21}. ID 120 shows visible signatures of interactions or mergers and as expected, it shows a higher value of asymmetry. We note that for ID-232, the asymmetry parameter in MIRI is nearly three times that of HST or NIRCam. This is because of the detection of additional (dust) clump towards the West of the galaxy, as shown in the bottom panel of Fig. \ref{fig:multiband_images}. Therefore, general trends in the average CAS values agree well with the findings in the literature, regardless of the wavelength. 

We then compare the CAS parameters derived from HSC i-band and HST/F814W images to assess how these measurements differ between lower-resolution ground-based imaging and the higher-resolution HST data. We choose I-band ($\sim$8300 \r{A}) in both cases, to ensure there is no wavelength dependence introduced in the CAS parameters. Table \ref{tab:cas_hsc_hst} summarises the CAS values obtained from HSC and HST images. Note that no PSF matching has been applied between the HSC ($\sim$0.6 arcsec) and HST ($\sim$0.09 arcsec) datasets. From the values in Table \ref{tab:cas_hsc_hst}, we find that the concentration index (C) generally decreases from HSC to HST images, which is likely due to the narrower PSF of HST. The asymmetry (A) shows a slight increase in some galaxies, but remains broadly consistent across the two datasets. Similarly, clumpiness (S) is also largely consistent between HSC and HST images. Overall, the CAS parameters derived from ground-based and space-based imaging appear broadly consistent.

We then compared the morphological parameters from the HST, NIRCam and MIRI images. However, before this comparison, we ensured that the images under analysis have the same spatial resolution. The MIRI images have three times worse spatial resolution ($\sim$0.28 arcsec) compared to HST and NIRCam images (both $\sim$0.09 arcsec). Therefore, as mentioned earlier, the HST and NIRCam images were brought to the resolution of the MIRI images, by convolving them with a Gaussian kernel. Figure \ref{fig:CAS_HST_NIRCam_MIRI} shows the CAS parameters as a function of wavelength (HST = 8333 \r{A}, NIRCam $\sim 2.7 ~\mu$m and MIRI $\sim 7.7 ~\mu$m) for each target. The CAS parameters in Fig. \ref{fig:CAS_HST_NIRCam_MIRI} have been normalised to the values of the HST images to better guide the relative changes in the parameters for each target. From this figure and from the values reported in Table \ref{tab:cas_miri_nircam}, it is clear that the morphological parameters change as a function of wavelength. Generally, the CAS parameters appear to increase as a function of wavelength in absolute terms, with MIRI images show the maximum dispersion or range in the CAS parameters compared to NIRCam or HST images.

We calculate the differences in the individual parameters, namely, change in concentration ($\Delta$C), asymmetry ($\Delta$A), clumpiness ($\Delta$S), M$_{20}$ ($\Delta$M$_{20}$) and Gini ($\Delta$Gini) for each target. These differences in the morphological parameters can also be visualised for individual targets via the plots in Figs. \ref{fig:DeltaCAS_plots} and \ref{fig:DeltaM20Gini_plots}. The top panels in Fig. \ref{fig:DeltaCAS_plots} show $\Delta$C (left panel), $\Delta$A (middle panel) and $\Delta$S (right panel) for different morphological types, classified visually (ETGs: E/S0 or LTGs: S). The lower panels in Fig. \ref{fig:DeltaCAS_plots}, on the other hand, show $\Delta$C, $\Delta$A and $\Delta$S as a function of stellar mass, M$_{\ast}$. Figure \ref{fig:DeltaM20Gini_plots} shows similar plots, but for M$_{20}$ and Gini parameters. The data points in these plots are further separated into the difference in each band: blue shows HST$-$NIRCam, orange shows NIRCam$-$MIRI and red shows HST$-$NIRCam. Only the absolute values in these changes are displayed in these plots.

On average, the morphological difference calculated from HST$-$NIRCam datasets show the least variation compared to NIRCam$-$MIRI and HST$-$MIRI. This is clear from the dashed blue, orange and red lines in Figs. \ref{fig:DeltaCAS_plots} and \ref{fig:DeltaM20Gini_plots}, which correspond to average values of the parameter differences. This is consistent with the trends observed in Fig. \ref{fig:CAS_HST_NIRCam_MIRI} and also while visually analysing images in Figs. \ref{fig:multiband_images}, \ref{fig:multiband_images_1} and \ref{fig:multiband_images_2}. For example, in ID-232 in the bottom panel in Fig. \ref{fig:multiband_images}, the clump towards the West edge of the galaxy is prominent in the MIRI image (right panel), but is otherwise absent in the HST and NIRCam images in the left and middle panels, respectively. The contrary is also true i.e, due to the depth of the NIRCam and HST images, some of the structures are more prominent or clearly visible in NIRCam and HST images instead. For example, the late-type morphology of ID-269 is evident in HST and NIRCam images in the bottom panel of Fig. \ref{fig:multiband_images_2}. However, the MIRI image depth is not sufficient to detect the extended structure. 

We do not find any trend between $\Delta$C, $\Delta$A and $\Delta$S with stellar mass. The M$_{20}$ parameter is relatively stable with stellar mass, as evident from the right panel in Fig. \ref{fig:DeltaM20Gini_plots}. The lack of a clear trend with either morphological type or stellar mass may be due to the relatively narrow range in stellar mass probed in the dwarf galaxies presented here. 

Figure \ref{fig:CAS_plot} shows the distribution of these morphological parameters in different parameter spaces, namely clumpiness versus asymmetry (left panel), asymmetry versus concentration (middle panel) and clumpiness versus concentration (right panel). These plots have usually been used to distinguish between early-type and late-type galaxies in the massive galaxy regime, however this may not be true in the dwarf regime \citep{lazar24}. To visualise where the dwarf galaxies presented in this paper lie with respect to the parent sample of dwarf galaxies from \citet{lazar24}, we also plot the CAS parameters for the parent sample, obtained from the non-parametric analysis in HSC data, in the same plots (shown as grey data points). We expect some differences between the location of the parent sample in \citet{lazar24} from the HST and JWST images presented here due to the differences in the PSF and wavelength changes, as already discussed earlier. This is indeed apparent from the plots in Fig. \ref{fig:CAS_plot}. In most cases, the CAS parameters derived from HST, NIRCam and MIRI images reside in similar location as the parent sample of dwarf galaxies from \citet{lazar24}. A similar trend is observed in the M$_{20}$ versus Gini coefficient parameter space in Fig. \ref{fig:M20_vs_Gini} i.e., the parameters derived from HST, NIRCam and MIRI largely follow a similar locii as the parent \citet{lazar24} sample. The most deviation in the CAS parameters or M$_{20}$ and Gini parameters compared to \citet{lazar24} data points is seen in the case of MIRI data, which, as we have already seen from Figs. \ref{fig:CAS_HST_NIRCam_MIRI}, \ref{fig:DeltaCAS_plots} and \ref{fig:DeltaM20Gini_plots}, is most likely due to contribution from dust emission/clumps (see also Sect. \ref{sect4.2.1}), which become apparent at mid-infrared wavelengths, compared to optical or near-infrared wavelengths.

\section{Discussion} \label{sect5}

We presented multi-wavelength imaging of nine dwarf galaxies from the COSMOS-Web survey, combining data from HST/ACS with JWST/NIRCam and MIRI. The high-sensitivity and depth of imaging surveys with new instruments aboard telescopes such as JWST, and even Euclid, is now allowing us to target larger samples of low-mass galaxies across a wide range of redshifts and in low-density environments \citep[see][]{marleau_euclid25, shields25} Two key findings emerge from the analysis presented in this paper: First, all dwarf galaxies in our sample (100\%) were detected with both NIRCam and MIRI. When comparing the observed fluxes to those predicted for low-mass galaxies in the TNG50 simulations, we find that the predicted mid-infrared fluxes from TNG50 are roughly consistent with the observed MIRI fluxes. Second, the morphological parameters of the dwarf galaxies exhibit a slight dependence on wavelength, with the most significant variations observed in the MIRI images compared to those from HST and NIRCam. The distinct appearance of the MIRI images possibly reflects the presence of dust lanes and clumps, which is also evident from the results of the SED modelling. In this section, we explore the implications of the analysis presented in this paper. 

Broadly speaking, the mock fluxes derived from TNG50 simulations are dependent on: (1) the feedback prescriptions and the physics invoked in the TNG50 simulation itself, and (2) the RT implementation that assumes certain characteristics of the ISM, such as dust grain composition, dust-to-gas ratio and metallicity dependence. We discuss here how the change in the nature of the feedback or in any of the assumptions in the post-processing might change the simulated fluxes. Within TNG50, galactic-scale winds from stellar feedback are sufficiently strong enough to disrupt the formation of disc or further accretion of gas onto host galaxies \citep[see][]{celiz25}. However, these winds do not couple hydrodynamically with the surrounding ISM until they reach regions with densities below the star formation threshold \citep[see][]{martin25}. Such a treatment of stellar feedback may not be sufficient to quench star formation on shorter timescales. On the other hand, AGN feedback in the radiatively inefficient regime is rather weak and does not have a strong impact on the ISM. Traditionally, AGN feedback is expected to play a crucial role in the evolution of massive galaxies. However, it has been shown recently that both AGN and stellar feedback may act simultaneously on dwarf galaxies \citep[e.g.,][]{koudmani22, sharma23, arjona-galvez24}. Therefore, the ISM content or composition can change significantly, depending whether a radiatively efficient AGN feedback model is included in simulations \citep[e.g.,][]{farcy25}.

In addition to feedback, RT applied to reproduce mock fluxes also includes several assumptions that may affect our results. This includes a change in dust-to-gas ratios, as higher dust content could result in a higher radiative efficiency and stronger outflows \citep[e.g.,][]{fabian08}. So, we first focus on one of the free parameters in the RT procedure, $f_{\rm dust}$, which represents the fraction of metallic gas confined in dust grains. 
As discussed in \citet{trcka22}, this parameter varies across different samples in the literature and there are indications that $f_{\rm dust}$ depends on the metallicity, stellar mass and/or redshift. The $f_{\rm dust}=0.2$ is used for comparisons here, which is based on a median value from a sub-sample of galaxies from DustPedia galaxies \citep[see][]{DeVis19}, which have a stellar mass in the range log M$_{\ast} = 8-11.5$ (the median stellar mass skewed towards higher stellar mass of $\sim 10^{9}$ M$_{\odot}$). The $f_{\rm dust}$ parameter shows a large scatter in the stellar mass range probed in this paper (Figure 11 in \citet{DeVis19}). The large scatter is also visible for low metallicity environments. We are unable to probe the dependence of $f_{\rm dust}$ on metallicity for this sample, as we do not have estimates for the latter. Dwarf galaxies are generally expected to have low dust content and low metallicity. Using lower values of dust to-metal ratio for the diffuse dust, or higher covering fractions or lower HII region compactness in the \texttt{MAPPINGS} modelling of young stars may have resulted in lower mock MIRI fluxes (see \citealt{Liang2021}). However, the good agreement between modelled and observed fluxes suggest that these modelling choices are relatively sensible.
The multi-wavelength images and SED models presented in this paper clearly show that some of the dwarf galaxies presented here are abundant in dust. The SED models shown in Figs. \ref{fig:sed_fits} and \ref{fig:sed_fits_appendix} also suggest excess contribution from diffuse dust and/or PAHs at mid-infrared wavelengths, compared to contribution from a stellar source modelled by a simple black body fit. In most galaxies, more than 50\% of the mid-infrared flux is estimated to be non-stellar origin, likely from warm or hot dust or PAHs, which is also evident from the best-fit SED models in Figs. \ref{fig:sed_fits} and \ref{fig:sed_fits_appendix}.

The MIRI images several dwarf galaxies show indications of dust lanes or clumps, when compared to the HST and NIRCam images. In massive early-type galaxies, such dust lanes are usually indicative of previous merger activity with low mass companion galaxies \citep[see][]{kaviraj12, davis15}, suggesting that these low mass dwarf galaxies might also have gone through a merger. Nearly 7\% of the massive ETGs show such dust features \citep[e.g.,][]{kaviraj10}, compared to 22\% fraction we find in the dwarf galaxies (2/9). The incidence of dust lanes in our dwarf sample (in two sources: ID-5 and ID-175) is in contrast with 1\% reported in \citet{lazar24}. The multi-wavelength data, especially the addition of MIRI imaging helps detect the spatial offsets, which may explain such differences between our work and \citet{lazar24}. The combination of HST+NIRCam+MIRI imaging of dwarf galaxies in low-density environments suggests that the contribution of dust grains at mid-infrared wavelengths can be significant in these low-mass galaxies. Rest-frame UV and/or optical spectroscopy would confirm if higher dust content would also correlate with higher gas-phase metallicity in these galaxies. The MIRI/F770W images also corresponds to the 7.7 $\mu$m PAH emission, which may be stronger in some of the dwarfs presented here, as suggested by the best-fit SED models of at least three dwarfs (ID-132, ID-175 and ID-232). If confirmed via mid-infrared spectroscopy, this would be in contrast with previous studies of Blue Compact Dwarfs (BCDs) or other nearby dwarf galaxies \citep[e.g.,][]{madden06, hunt10}, that show a lack of PAH emission.

Our results are similar to the findings of \citet{wu06}, who show the presence of PAH emission in BCDs using Spitzer/IRS spectroscopy \citep[see also][]{hunt10}. Similarly, nearby dwarf galaxies (at distances $<30$ Mpc) in the Spitzer IRS Nearby Galaxies (SINGS) survey also reported the the presence of dust emission and that the dust tends to be clumpy \citep[e.g.,][]{dale07}. However, other studies have also reported weakened PAHs and lower dust content in nearby dwarf galaxies \citep[see review by][]{galliano18}. The typically low metallicities and weak PAH emission observed in dwarf galaxies may arise because PAHs are readily destroyed by intense UV radiation, while the small spatial scales of dwarf systems allow supernova-driven shocks to disperse PAHs globally, resulting in the characteristic PAH deficiency in low-metallicity galaxies \citep[e.g.,][]{remy-ruyer15}. We note that the studies described above largely focus on nearby dwarf galaxies or a special class of dwarf galaxies (BCDs), which may be biased towards higher SFRs and therefore, may not be representative of the broader dwarf population in low-density environments, such as the one discussed in this paper. Future mid-infrared spectroscopy with JWST/MIRI-MRS can provide strong constraints on the PAH composition in these dwarf galaxies. 

Finally, the TNG50 galaxies are resolved into few gas elements at the lowest stellar masses; this may have implications for the robustness of the RT post processing. This is one limitation of the mock fluxes, in addition to other choices involved in the modelling of diffuse dust and young stars.  
Nevertheless, in spite of these caveats, from the comparisons shown in Fig. \ref{fig:sim_comparison}, we conclude that the 100\% detection rate of dwarf galaxies in the COSMOS-Web survey is not surprising and that the observed fluxes are approximately within the flux range expected from TNG50.

\section{Conclusions and Final remarks} \label{sect6}

We present high resolution multi-wavelength HST and JWST imaging observations of a sample of nine dwarf galaxies ($10^{8} <$ M$_{\ast}$/M$_{\odot}< 10^{9}$) selected from the COSMOS-Web survey. Here we summarise the results of the analysis presented here: 

\begin{itemize}
\item All dwarf galaxies in the sample are detected with both NIRCam and MIRI, corresponding to a 100\% detection rate. These galaxies exhibit a broad range of colours, star formation rates (SFRs), and stellar masses (M$_\ast$), with an approximately equal split between early-type and late-type morphologies. This suggests that the sample is broadly representative of dwarf galaxies found in low-density environments.

\item The observed NIRCam and MIRI fluxes align reasonably well with predictions from low-mass galaxies in the TNG50 simulation. This consistency implies that the feedback models and radiative transfer parameters used in post-processing are suitable for modelling such low-mass systems. Follow-up UV/optical spectroscopy will be essential for constraining the ionisation parameter/structure, metallicity, metallicity gradient, star formation rate and extinction within these galaxies.

\item The morphological parameters—CAS, M$_{20}$, and Gini—derived from HST, NIRCam, and MIRI data are generally consistent with those from the parent sample of \citet{lazar24}. Deviations are primarily seen in the MIRI-derived values, likely due to the influence of dust and/or the shallower depth of MIRI observations compared to NIRCam and HST. Owing to the diversity in the ISM properties of dwarf galaxies, we do not find a clear trend between these morphological parameters and the stellar mass or galaxy type. Surveys with future facilities such as LSST and ROMAN space telescope will be crucial to further explore the morphological diversity in a statistical sample of dwarf galaxies. 

\item Mid-infrared MIRI observations strongly indicate the presence of dust in these dwarf galaxies, either in the form of dust clumps or dust lanes. Follow-up spectroscopy in near and mid-infrared wavelengths using NIRSpec and MIRI-MRS will provide crucial constraints on the contribution to mid-infrared fluxes from PAHs, which will further provide detailed insights into the dust grain composition in these dwarf galaxies. MRS spectroscopy will also provide access to rotational transitions of hydrogen molecule and other ionisation species, which will aid in constructing diagnostic diagrams to infer the molecular gas content and source of ionisation in these dwarf galaxies. 
\end{itemize}

\section*{Acknowledgements}
We thank the anonymous referee for their constructive comments and suggestions. We thank M. Shuntov for providing us with the SED fits. SK, IL and AEW acknowledge support from the STFC (grant numbers ST/Y001257/1 and ST/X001318/1). SK acknowledges a Senior Research Fellowship from Worcester College Oxford. IL and BB acknowledge PhD studentships funded by the Centre for Astrophysics Research at the University of Hertfordshire. RKC is grateful for support from the Leverhulme Trust via the Leverhulme Early Career Fellowship. SK has been supported by a Junior Research Fellowship from St Catharine's College, Cambridge and a Research Fellowship from the Royal Commission for the Exhibition of 1851. This research is based on observations made with the NASA/ESA Hubble Space Telescope obtained from the Space Telescope Science Institute, which is operated by the Association of Universities for Research in Astronomy, Inc., under NASA contract NAS 5–26555. These observations are associated with program 9822 and 10092. This work is based [in part] on observations made with the NASA/ESA/CSA James Webb Space Telescope. The data were obtained from the Mikulski Archive for Space Telescopes at the Space Telescope Science Institute, which is operated by the Association of Universities for Research in Astronomy, Inc., under NASA contract NAS 5-03127 for JWST. These observations are associated with program 1727. 

\section*{Data Availability}
The data used in this paper are available from the MAST portal (mast.stsci.edu) from the HST programme IDS 9822 and 10092 and JWST programme ID 1727. 



\bibliographystyle{mnras}
\bibliography{reference} 




\appendix

\section{Appendix} \label{sect:appendix}

\begin{figure*}
\centering
\includegraphics[width=0.8\textwidth]{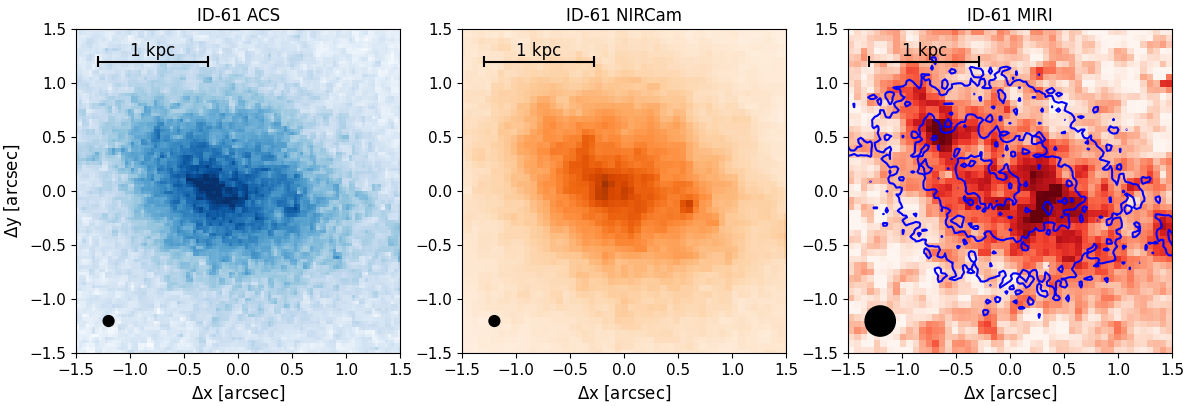}
\includegraphics[width=0.8\textwidth]{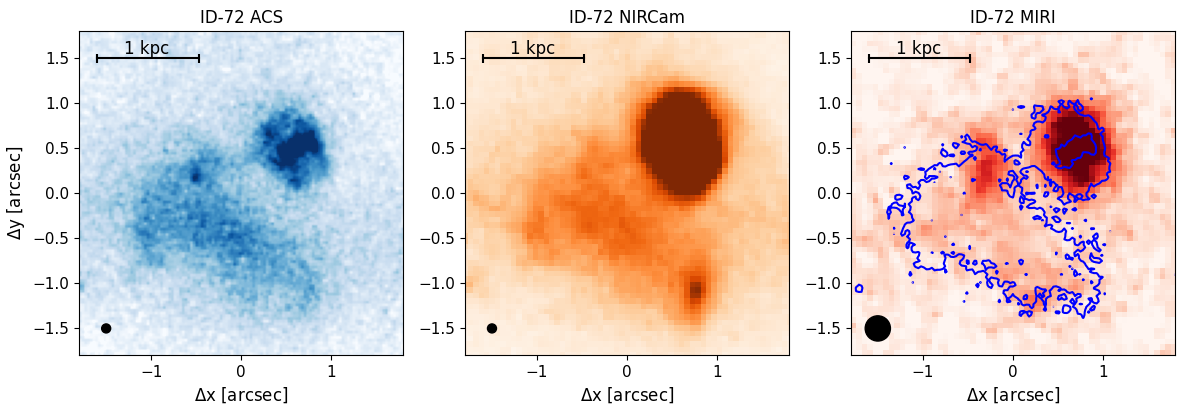}
\includegraphics[width=0.8\textwidth]{HST_JWST_imgs_120.png}
\includegraphics[width=0.8\textwidth]{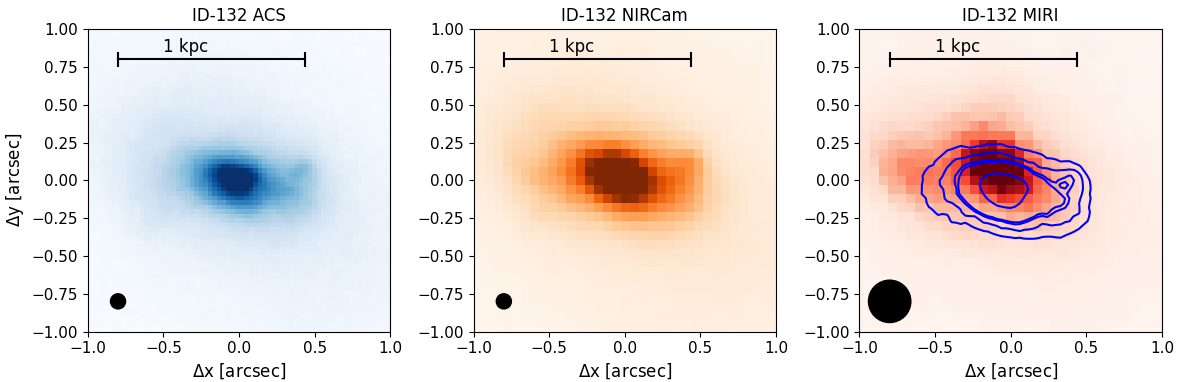}
\caption{From top to bottom: ID 61, 72, 120 and 132}
\label{fig:multiband_images_1}
\end{figure*}

\begin{figure*}
\centering

\includegraphics[width=0.8\textwidth]{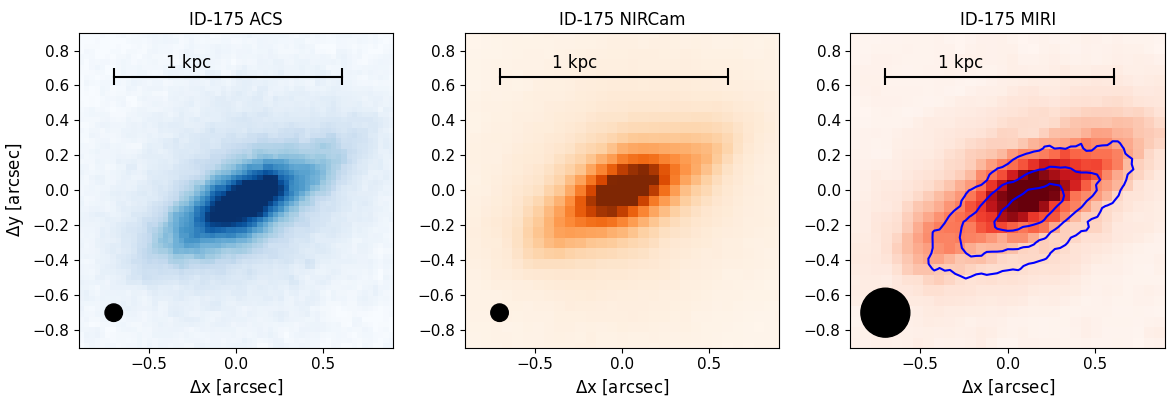}
\includegraphics[width=0.8\textwidth]{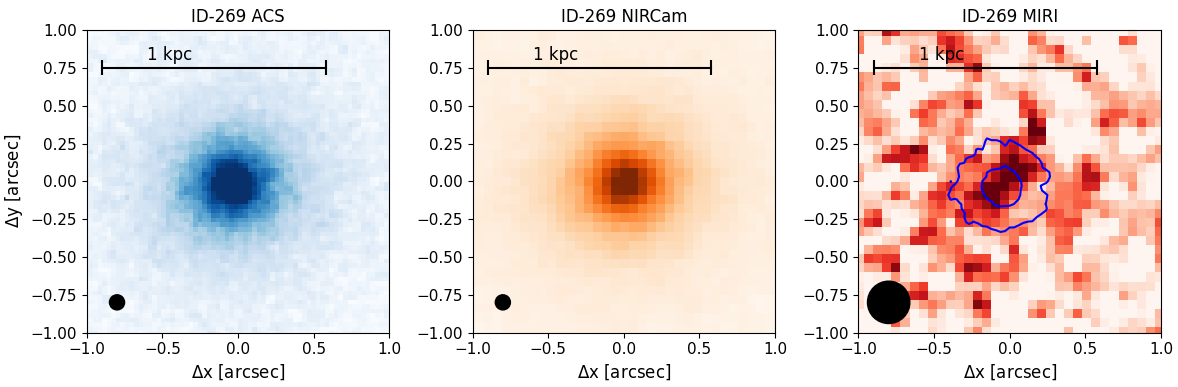}
\caption{From top to bottom: ID 175 and 269}
\label{fig:multiband_images_2}
\end{figure*}

\begin{figure*}
\centering
\includegraphics[width=0.48\linewidth]{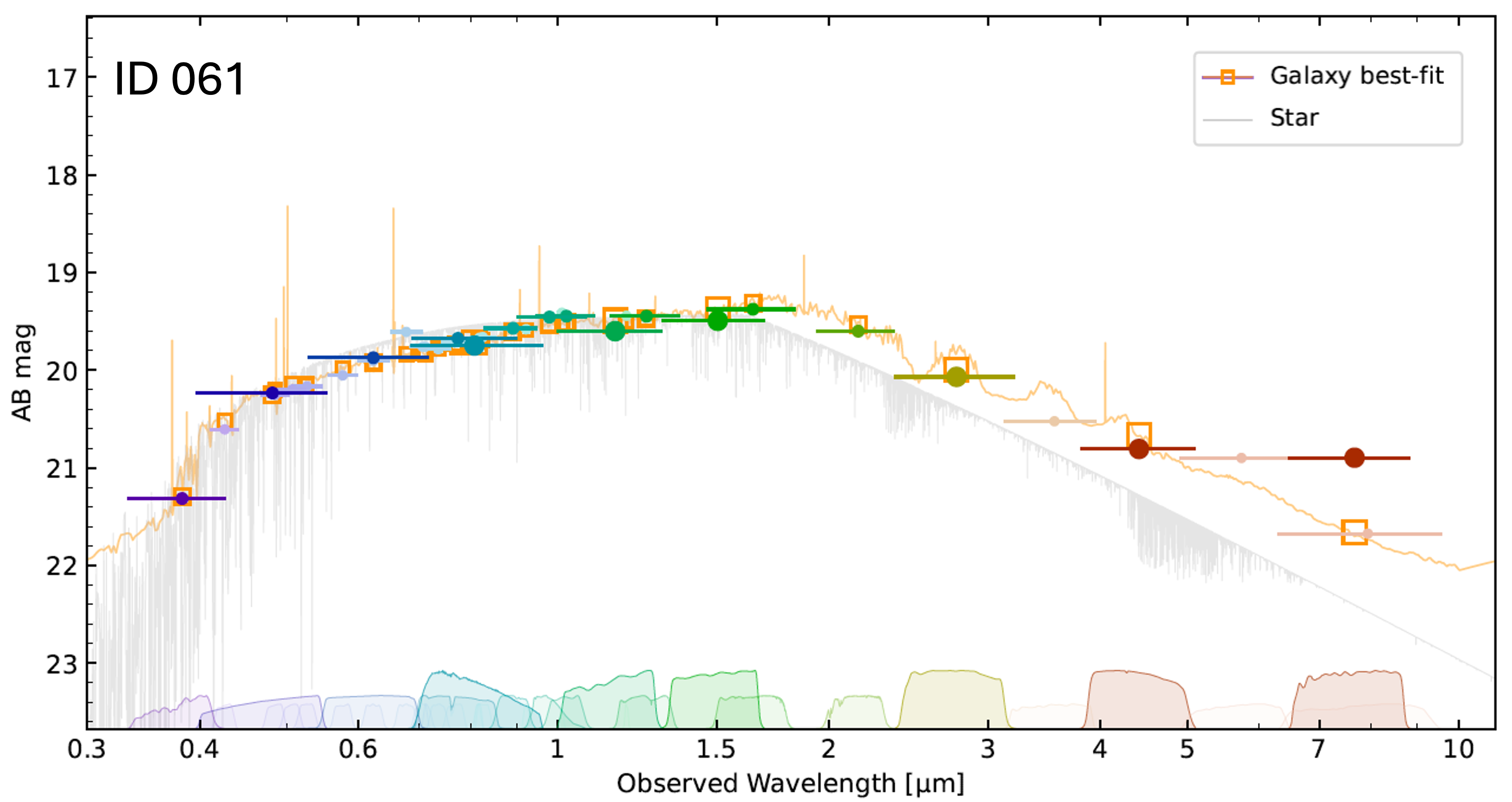}
\includegraphics[width=0.48\linewidth]{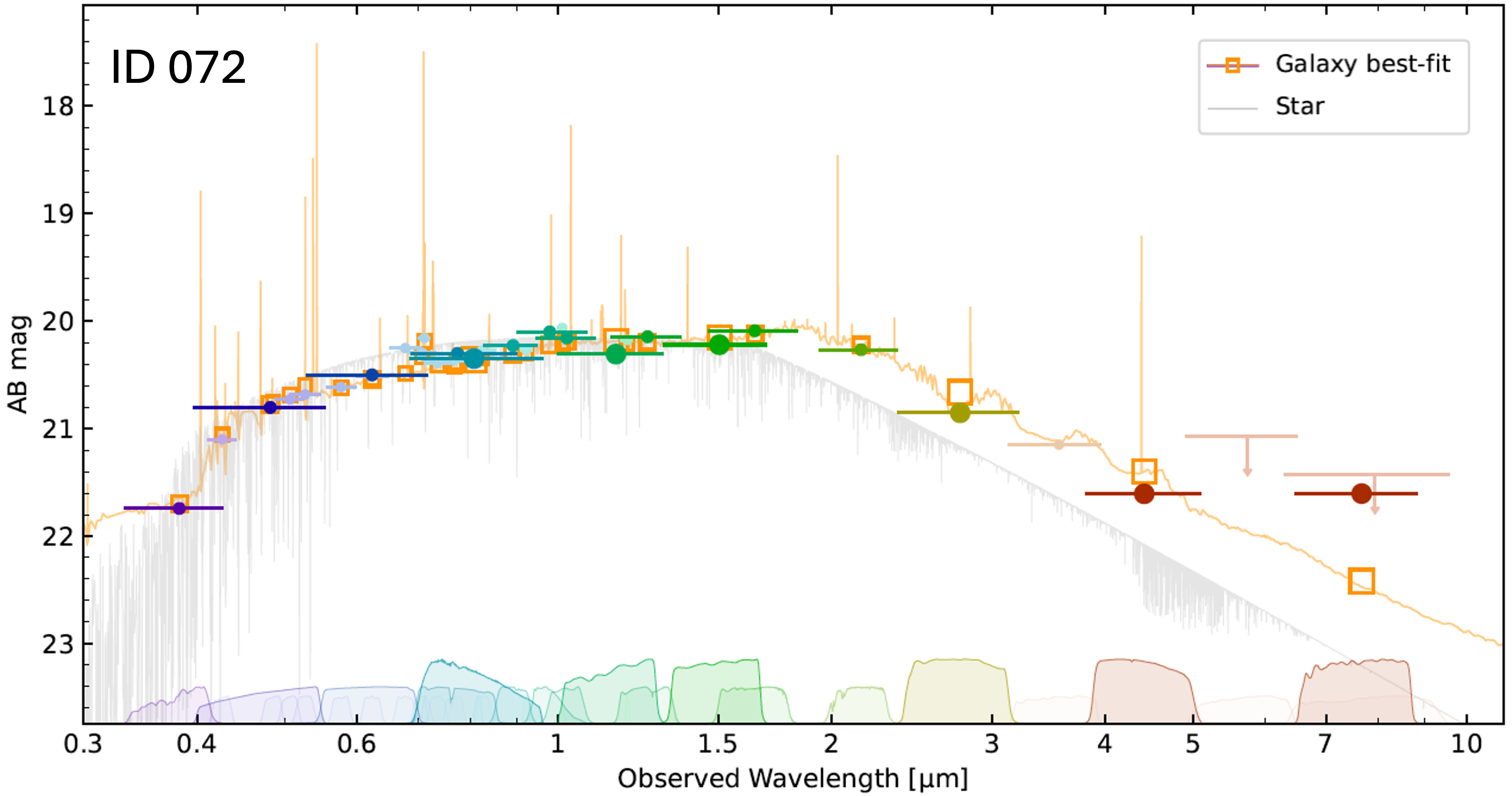}\\
\includegraphics[width=0.48\linewidth]{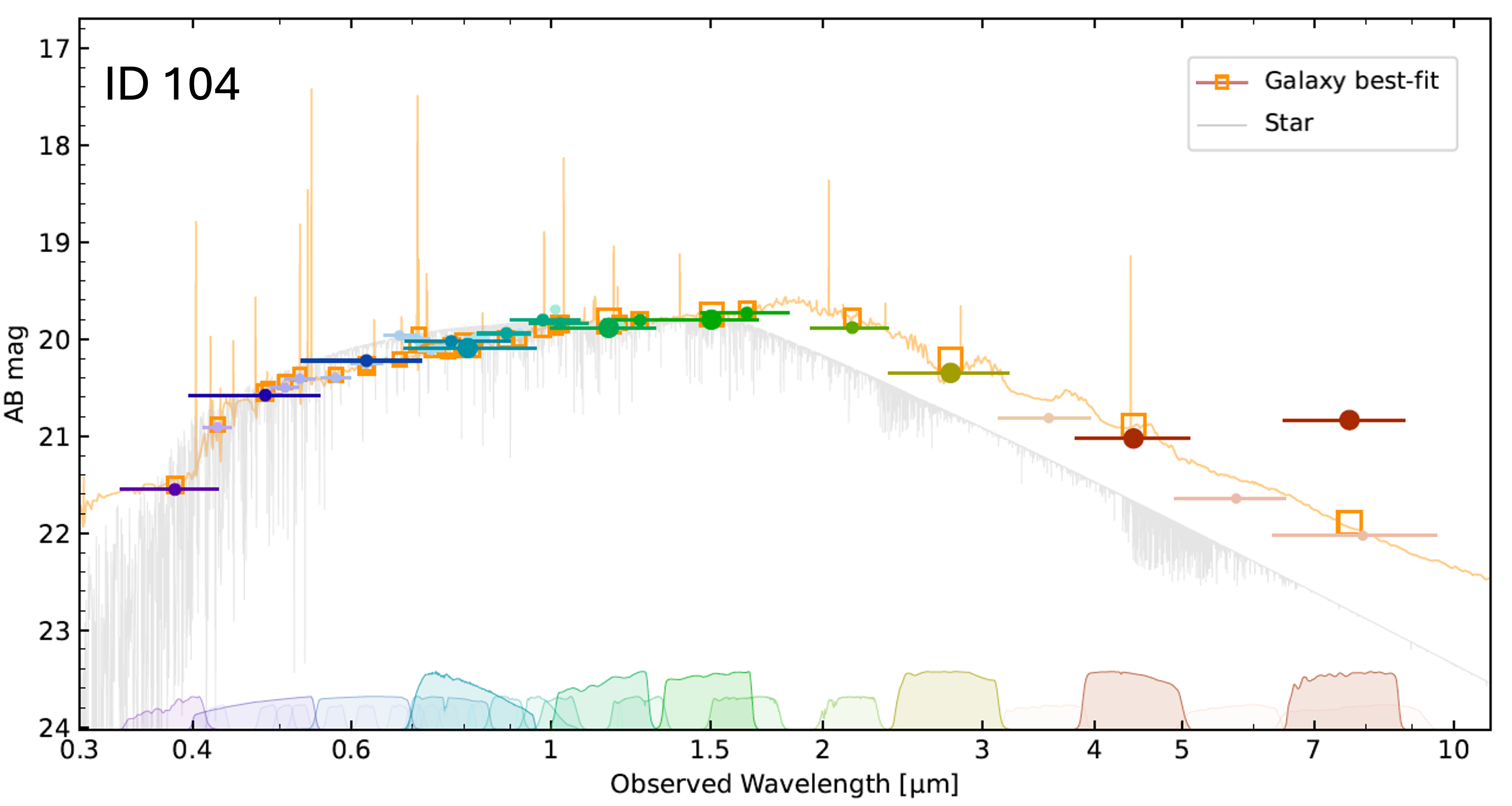}
\includegraphics[width=0.48\linewidth]{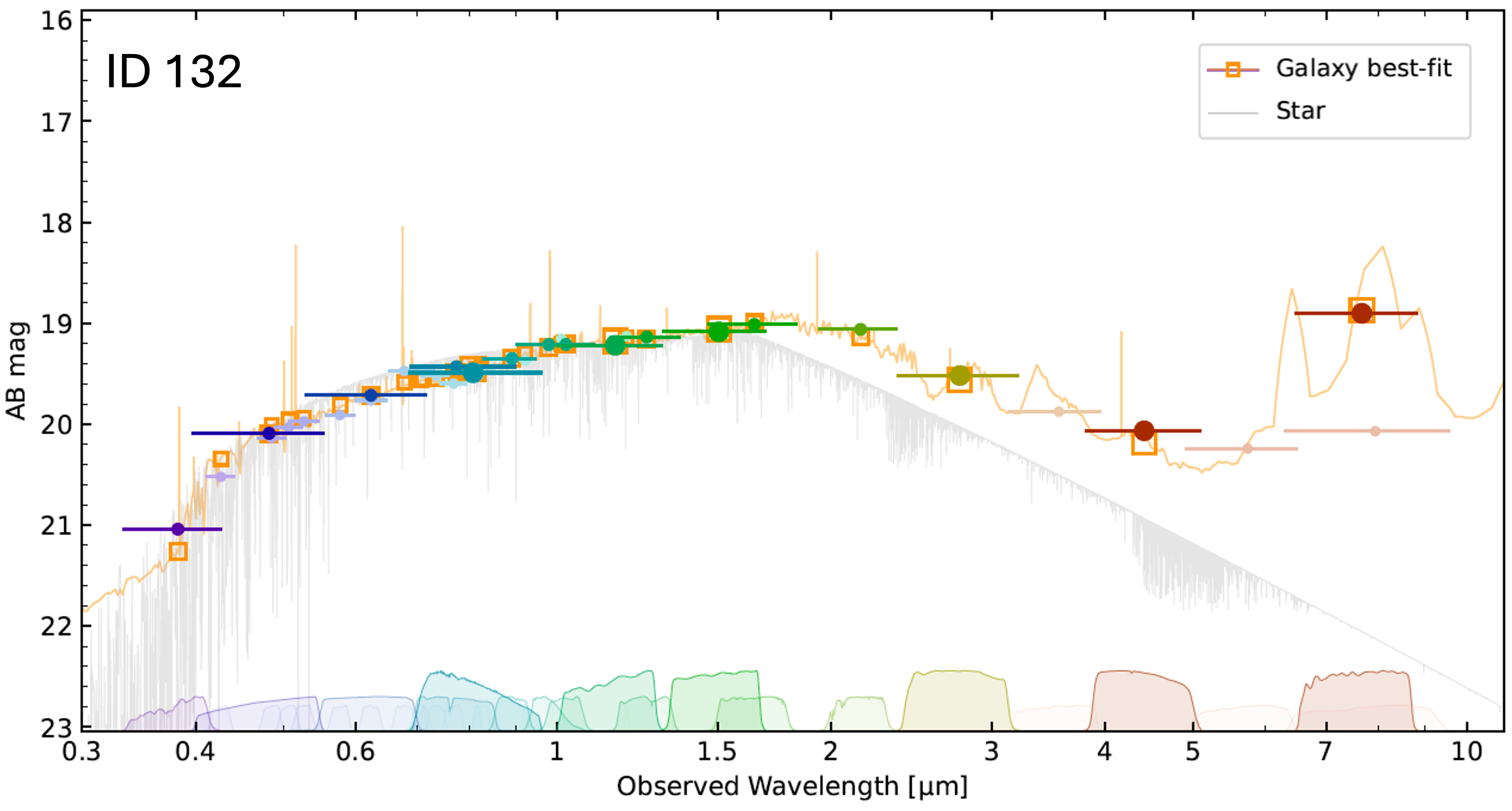}\\
\includegraphics[width=0.48\linewidth]{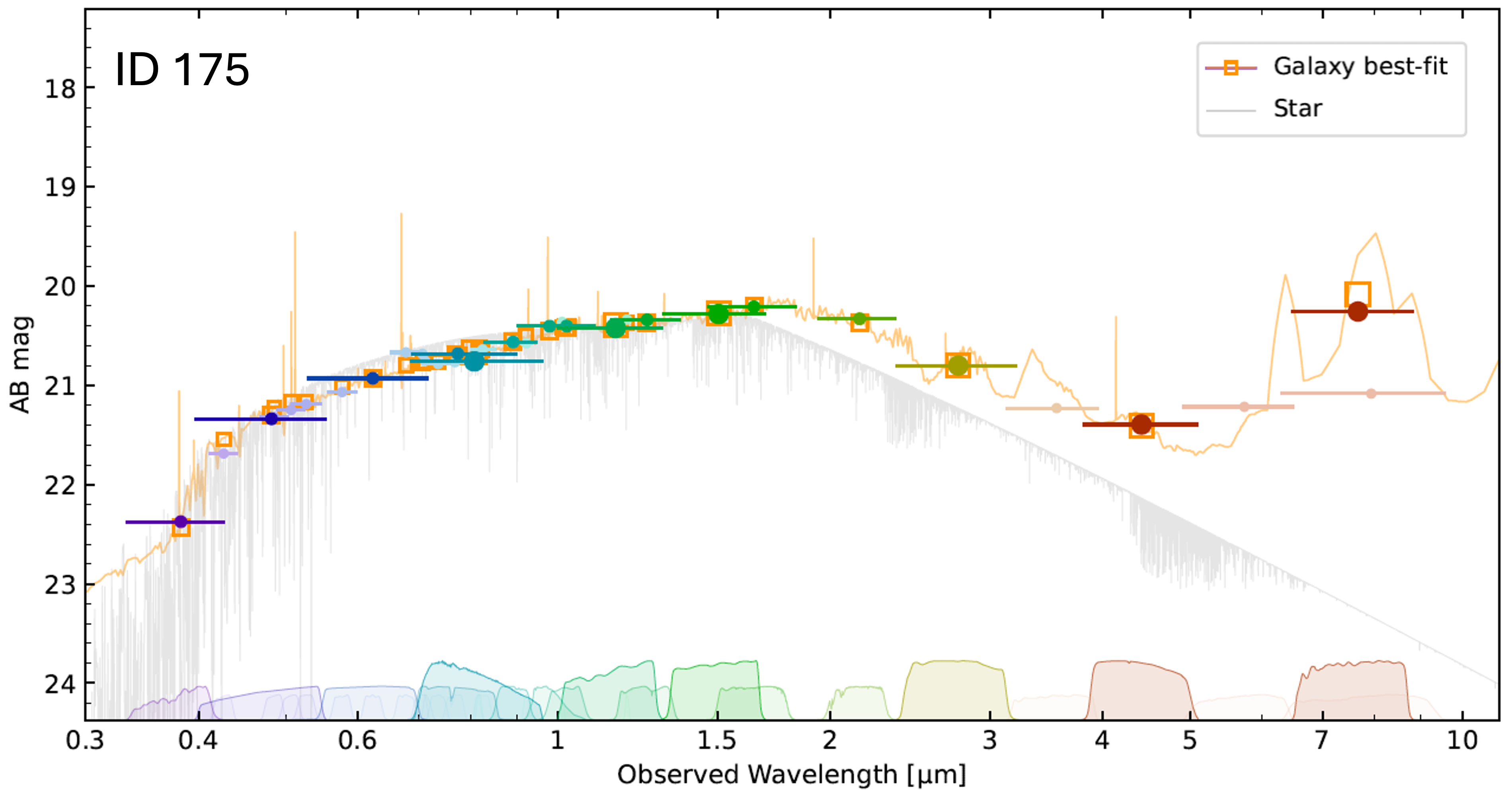}
\includegraphics[width=0.48\linewidth]{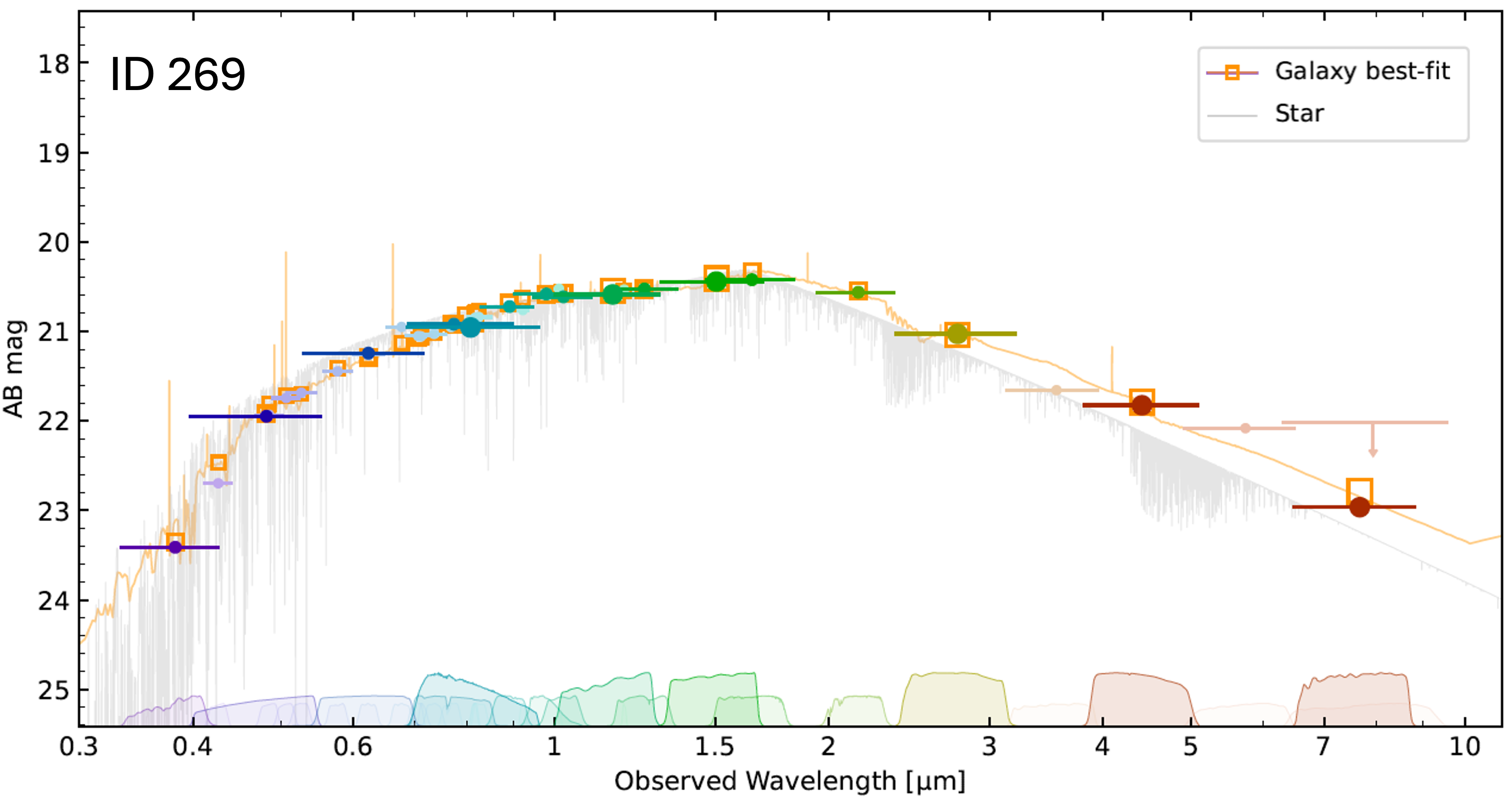}\\
\caption{SED fits of the rest of the targets in this paper. The colour scheme is the same as Fig. \ref{fig:sed_fits}.}
\label{fig:sed_fits_appendix}
\end{figure*}

\begin{table*}
\centering
\begin{tabular}{l|cccccc|cccccc|cccccc|}
ID & \multicolumn{5}{c}{HST/ACS} & \multicolumn{5}{c}{NIRCam} & \multicolumn{5}{c}{MIRI}\\
\hline
& $n$ & C & A & S & M$_{20}$ & Gini & $n$ & C & A & S & M$_{20}$ & Gini & $n$ & C & A & S & M$_{20}$ & Gini\\
\hline\hline
5 & 1.0 & 2.90 & 0.03 & 0.004 & -1.80 & 0.52 & 1.0 & 2.90 & 0.01 & 0.02 & -1.81 & 0.52 & 0.9 & 2.58 & -0.02 & 0.05 & -1.48 & 0.43\\
61 & 1.0 & 2.54 & -0.02 & -0.03 & -1.71 & 0.44 & 1.0 & 2.77 & -0.04 & -0.01 & -1.68 & 0.50 & 1.0 & 2.67 & 0.10 & -0.06 & -1.36 & 0.48\\
72 & 0.6 & 1.78 & 0.0 & 0.0 & -1.27 & 0.23 & 0.5 & 2.03 & 0.00 & 0.04 & -1.27 & 0.38 & 0.6 & 2.46 & 0.29 & 0.11 & -1.32 & 0.45\\
104 & 1.1 & 2.92 & 0.03 & 0.01 & -1.79 & 0.52 & 1.0 & 2.93 & 0.03 & 0.01 & -1.80 & 0.51 & 0.5 & 2.46 & 0.49 & 0.14 & -1.23 & 0.51\\
120 & 0.7 & 2.26 & 0.30 & 0.02 & -1.23 & 0.43 & 0.7 & 2.45 & 0.28 & 0.03 & -1.30 & 0.46 & 0.8 & 1.54 & 0.60 & 0.05 & -0.98 & 0.51\\
132 & 1.1 & 2.86 & 0.09 & 0.004 & -1.80 & 0.52 & 1.1 & 2.94 & 0.10 & 0.04 & -1.84 & 0.52 & 1.1 & 2.96 & 0.27 & 0.02 & -1.70 & 0.51\\
175 & 0.9 & 2.74 & 0.02 & 0.02 & -1.75 & 0.50 & 0.8 & 2.71 & 0.02 & 0.003 & -1.74 & 0.50 & 0.8 & 2.82 & 0.07 & 0.02 & -1.78 & 0.48\\
232 & 0.8 & 2.66 & 0.11 & 0.02 & -1.55 & 0.44 & 0.9 & 3.08 & 0.16 & 0.04 & -1.67 & 0.52 & 1.0 & 3.19 & 0.36 & 0.05 & -1.09 & 0.57\\
269 & 1.2 & 3.07 & 0.03 & -0.003 & -1.89 & 0.53 & 1.1 & 3.03 & 0.03 & -0.002 & -1.86 & 0.53 & 1.2 & 2.32 & -0.28 & -0.07 & -0.95 & 0.43\\
\hline
Avg & 0.9 & 2.67 & 0.10 & 0.006 & -1.65 & 0.48 & 0.9 & 2.79 & 0.14 & 0.019 & -1.67 & 0.51 & 1.0 & 2.61 & 0.20 & 0.023 & -1.37 & 0.50\\
\hline
\end{tabular}
\caption{Parametric and non-parametric morphological analysis of the dwarf galaxies presented in this paper. The notations are the same as in Table \ref{tab:cas_hsc_hst}.}
\label{tab:cas_miri_nircam}
\end{table*}


\bsp	
\label{lastpage}
\end{document}